\documentclass[%
 jmp,
 amsmath,amssymb,
reprint, onecolumn 
]{revtex4-1}

\usepackage{graphicx}
\usepackage{dcolumn}
\usepackage{bm}

\usepackage[utf8]{inputenc}
\usepackage[T1]{fontenc}
\usepackage{mathptmx}
\usepackage{etoolbox}

\makeatletter
\def\@email#1#2{%
 \endgroup
 \patchcmd{\titleblock@produce}
  {\frontmatter@RRAPformat}
  {\frontmatter@RRAPformat{\produce@RRAP{*#1\href{mailto:#2}{#2}}}\frontmatter@RRAPformat}
  {}{}
}%
\makeatother
\begin{document}

\preprint{AIP/123-QED}

\title[Sample title]{On a discrete version of the position-momentum commutation relation}
\author{Nicolae Cotfas}
\email{nicolae.cotfas@g.unibuc.ro}
\affiliation{ 
Faculty of Physics, University of Bucharest, P.O.Box MG-11, Bucharest, 077125, Romania
}%

 \homepage{https://unibuc.ro/user/nicolae.cotfas/}

\date{\today}

\begin{abstract}
The usual position-momentum commutation relation plays a fundamental role in the
mathematical description of continuous-variable quantum systems.
In the case of a qudit described by a Hilbert space of a high enough dimension, there exists
a discrete version of the position-momentum commutation relation satisfied with approximation of a large part of the pure quantum states. Our purpose is to explore in more
details the set of these states. We show that it contains a family of discrete-variable
Gaussian states depending on a continuous parameter and certain discrete coherent states. 
It also contains various discrete-variable versions of the 
Hermite-Gauss states, defined either as eigenstates of certain discrete versions of the
harmonic oscillator Hamiltonian or generated by using a discrete version of the
creation or annihilation operator. The presented results suggest a possible way to experimentally obtain some of the investigated states. 
\end{abstract}

\maketitle

\section{Introduction}

The quantum systems with finite-dimensional Hilbert space 
\cite{Schwinger60,Vourdas04,Cotfas11,Cotfas23,Cotfas24}
play an important role in the investigation of quantum mechanics 
foundations, in Quantum Information, and in other applications.
The mathematical description of these discrete-variable 
systems is obtained by following, as much as possible, 
the analogy with the continuous-variable quantum systems.
In certain cases, the discrete-variable versions have 
properties very similar to those concerning the continuous
systems, but this does not happen in all the cases.

In quantum mechanics, a pair $A$ and $B$ of Hermitian operators on a 
Hilbert space is {\it canonical}  if they satify the relation $[A,B]\!=\!\mathrm{i}\hbar $, where $[A,B]\!=\!AB-\!BA$, $h\!=\!2\pi \hbar$ is the Plank constant, and $\mathrm{i}\hbar\equiv\mathrm{i}\hbar I$ is the identity operator multiplied by $\mathrm{i}\hbar$.
In the finite-dimensional case, such a pair does not exist because the trace of
the two sides of $[A,B]\!=\!\mathrm{i}\hbar $ is different.
In \cite{Farrales25}, the authors show that certain canonical pairs can be obtained by
relaxing the definition, namely by imposing the relation  $[A,B]\!=\!\mathrm{i}\hbar $ to be satisfied only on a proper subspace of the Hilbert space. Some references concerning possible applications of this new definition are presented in \cite{Farrales25}.

Our approach is different. We show that a discrete version of the position-momentum commutation relation $[\bar{q},\bar{p}]\!=\!\mathrm{i}\hbar $ is satisfied with an approximation in a proper subspace if the dimension of the Hilbert space is high enough.
Among the elements of such a subspace there are some of the most important states
used in the mathematical description of qudits \cite{Mehta87,Barker,Nikiforov,Vilenkin}.

The presented results may be used as a starting point of new mathematical formalisms
involving only certain particular states of a qudit,  by following, for example,
the analogy with the theory of continuous-variable Gaussian states used in Quantum Information \cite{Ferraro05,Wang07,Adesso14}.
%
%
\section{Quantum systems with finite-dimensional Hilbert space}
A spinless particle with mass, in one dimension,  is usually
described by using the Hilbert space 
\begin{equation}
L^2(\mathbb{R})\!=\!\overline{\left\{ 
\Psi\!:\!\mathbb{R}\!\rightarrow\!\mathbb{C}\ \left| \ 
\int_{-\infty}^\infty |\Psi(q)|^2\, dq\!<\!\infty\right.\right\}}
\end{equation}
(closure of the space of square integrable functions) with 
\begin{equation}
\left\langle \Phi,\Psi\right\rangle\!=\!
\int_{-\infty}^\infty \overline{\Phi(q)}\, \Psi(q)\, dq.
\end{equation}
In order to have a simple analogy, we shall consider only
the case of a discrete-variable quantum system described 
by an odd-dimensional ($d\!=\!2s\!+\!1$) Hilbert space 
$\mathcal{H}$  regarded as a space of functions, namely
\begin{equation}
\mathcal{H}\!=\!\{\psi\!:\!\{-\!s,\!-s\!+\!1,...,s\!-\!1,s\}\!\rightarrow \!\mathbb{C}\,\}
\end{equation}
with the inner product
\begin{equation}
\langle \varphi|\psi\rangle \!=\!\sum\limits_{n=-s}^s\overline{\varphi(n)}\,\psi(n)
\end{equation}
and the norm
\begin{equation}
||\psi||\!=\!\sqrt{\langle \psi|\psi\rangle} \!=\!\sqrt{\sum\limits_{n=-s}^s |\psi(n)|^2}.
\end{equation}
The set $\{-\!s,\!-s\!+\!1,...,s\!-\!1,s\}$ contains one and only one representative of any class of remainders modulo $d$. Therefore, we can identify it with $\mathbb{Z}_d$.
The discrete $\delta$ functions $\delta_{-s}$, $\delta_{-s+1}$,  ... , $\delta_{s}$,  \ where $\delta_m\!:\!\{-\!s,\!-s\!+\!1,...,s\}\!\rightarrow\!\mathbb{C},$
\begin{equation}
\begin{array}{l}
\delta_m(n)\!=\!\delta_{mn}\!=\!\left\{ \!\!\begin{array}{ccc}
1 & \mbox{for} & n\!=\!m,\\
0 & \mbox{for} & n\!\neq\!m,
\end{array}\right.
\end{array}
\end{equation}
form an orthonormal basis of $\mathcal{H}$.
By using Dirac notation $|m\rangle $ instead of $|\delta_m\rangle$, we have
\begin{equation}
\langle n|m\rangle\!=\!\delta_{nm},\quad \mbox{and}\quad \sum_{m=-s}^s|m\rangle\langle m|\!=\!\mathbb{I},
\end{equation}
where $\mathbb{I}\!:\!\mathcal{H}\!\rightarrow \!\mathcal{H},\ \mathbb{I}\psi\!=\!\psi$, is the identity operator.
\subsection{Discrete Fourier transform}
The usual Fourier transform \ $\Psi\!\mapsto F\Psi,$
\begin{equation}
\begin{array}{l}
(F\Psi)(p)\!=\!\frac{1}{\sqrt{2\pi \hbar}}\int\limits_{-\infty}^\infty \mathrm{e}^{-\frac{\mathrm{i}}{\hbar}pq}\, \Psi(q)\, dq\!=\!\frac{1}{\sqrt{h}}\int\limits_{-\infty}^\infty \mathrm{e}^{-\frac{2\pi \mathrm{i}}{h}pq}\, \Psi(q)\, dq, 
\end{array}
\end{equation}
is a unitary transform. Its inverse is the adjoint transform \ 
$\Psi\!\mapsto F^\dag\Psi,$
\begin{equation}
\begin{array}{l}
(F^\dag\Psi)(p)\!=\!\frac{1}{\sqrt{h}}\int\limits_{-\infty}^\infty\mathrm{e}^{\frac{2\pi \mathrm{i}}{h}pq}\, \Psi(q)\, dq, 
\end{array}
\end{equation}
The corresponding discrete-variable version \ $\mathbf{F}\!:\!\mathcal{H}\!\rightarrow\!\mathcal{H}\!:\!\psi \mapsto \mathbf{F}\psi,$ 
\begin{equation}
(\mathbf{F}\psi)(k)\!=\!\mbox{\small $\frac{1}{\sqrt{d}}$}\!\sum\limits_{n=-s}^s \mathrm{e}^{-\frac{2\pi \mathrm{i}}{d}kn}\, \psi(n),
\end{equation}
is also a unitary transform \cite{Vourdas04,Mehta87}.
Its inverse is the adjoint \ $\mathbf{F}^\dag\!:\!\mathcal{H}\!\rightarrow\!\mathcal{H}\!:\!\psi \mapsto \mathbf{F}^\dag\psi,$
\begin{equation}
(\mathbf{F}^\dag\psi)(k)\!=\!\mbox{\small $\frac{1}{\sqrt{d}}$}\!\sum\limits_{n=-s}^s \mathrm{e}^{\frac{2\pi \mathrm{i}}{d}kn}\, \psi(n).
\end{equation}
Particularly, \ $\langle n|\mathbf{F}|k\rangle\!=\!(\mathbf{F}\delta_k)(n)\!=\!\mbox{\small $\frac{1}{\sqrt{d}}$}\mathrm{e}^{-\frac{2\pi \mathrm{i}}{d}kn}$, \ 
$\langle n|\mathbf{F}^\dag|k\rangle\!=\!(\mathbf{F}^\dag\delta_k)(n)\!=\!\mbox{\small $\frac{1}{\sqrt{d}}$}\mathrm{e}^{\frac{2\pi \mathrm{i}}{d}kn}$, \ and
\begin{equation}
\begin{array}{l}
\mathbf{F}\!=\!\mathbb{I}\mathbf{F}\mathbb{I}\!=\!
\sum_{n,k=-s}^s|n\rangle\langle n|\mathbf{F}|k\rangle\langle k|\!=\!
\mbox{\small $\frac{1}{\sqrt{d}}$}\!\sum\limits_{n,k=-s}^s \mathrm{e}^{-\frac{2\pi \mathrm{i}}{d}kn}\, |n\rangle\langle k|,\\[1mm]
\mathbf{F}^\dag\!=\!\mathbb{I}\mathbf{F}^\dag\mathbb{I}\!=\!
\sum_{n,k=-s}^s|n\rangle\langle n|\mathbf{F}^\dag|k\rangle\langle k|\!=\!
\mbox{\small $\frac{1}{\sqrt{d}}$}\!\sum\limits_{n,k=-s}^s \mathrm{e}^{\frac{2\pi \mathrm{i}}{d}kn}\, |n\rangle\langle k|.
\end{array}
\end{equation}
\subsection{Discrete version of the position and momentum operators}
The position operator \ $\hat{q}\!:\!D_q\!\subset\!L^2(\mathbb{R})\!\rightarrow \!L^2(\mathbb{R})\!:\!\Psi\!\mapsto \hat{q}\, \Psi,$
\begin{equation}
\begin{array}{l}
(\hat{q}\, \Psi)(q)\!=\!q\, \Psi(q), 
\end{array}
\end{equation}
whose domain of definition is 
\begin{equation}
D_q\!=\!\left\{ \, \Psi \!\in\!L^2(\mathbb{R})\ \left| 
\ 
\int_{-\infty}^\infty |q\,\Psi(q)|^2\, dq\!<\!\infty\right.\right\},
\end{equation}
admits \cite{Schwinger60,Vourdas04} the discrete-variable version \ 
$\hat{\mathbf{q}}:\mathcal{H}\!\rightarrow\!\mathcal{H}\!:\!\psi \mapsto\hat{\mathbf{q}}\psi,$
\begin{equation}
(\hat{\mathbf{q}}\psi)(n)\!=\!n\, \psi(n).
\end{equation}
The basis $\{|n\rangle\}_{n=-s}^s$ is an eigenbasis for $\hat{\mathbf{q}}$, and $\hat{\mathbf{q}}$ admits the spectral decomposition
\begin{equation}
\hat{\mathbf{q}}\!=\!\sum\limits_{n=-s}^s n|n\rangle\langle n|.
\end{equation}

The momentum operator \ $\hat{p}\!:\!D_p\!\subset\!L^2(\mathbb{R})\!\rightarrow \!L^2(\mathbb{R})\!:\!\Psi\!\mapsto \hat{p}\, \Psi,$
\begin{equation}
\begin{array}{l}
\hat{p}\, \Psi\!=\!-\mathrm{i}\hbar \frac{\mathrm{d}}{\mathrm{d}q}\Psi, 
\end{array}
\end{equation}
whose domain of definition is 
\begin{equation}
D_p\!=\!\left\{ \, \Psi \!\in\!L^2(\mathbb{R})\ \left| 
\ 
\int_{-\infty}^\infty |\Psi'(q)|^2\, dq\!<\!\infty\right.\right\},
\end{equation}
satisfies the relation
\begin{equation}
\hat{p}\!=\!F^\dag \hat{q}F.
\end{equation}
The discrete-variable version \cite{Schwinger60,Vourdas04} of momentum is defined as \ $\hat{\mathbf{p}}\!:\!\mathcal{H}\!\rightarrow\!\mathcal{H},$
\begin{equation}
\hat{\mathbf{p}}\!=\!\mathbf{F}^\dag \hat{\mathbf{q}}\mathbf{F}.
\end{equation}
Explicitly,
\begin{equation}
\begin{array}{l}
(\hat{\mathbf{p}}\psi)(n)\!=\!\frac{1}{d}\sum\limits_{k=-s}^s\sum\limits_{m=-s}^s k \, \mathrm{e}^{\frac{2\pi \mathrm{i}}{d}k(n-m) }\, \psi(m),
\end{array}
\end{equation}
and $\hat{\mathbf{p}}$ admits the spectral decomposition
\begin{equation}
\hat{\mathbf{p}}\!=\!\sum\limits_{n=-s}^s n\, \mathbf{F}^\dag |n\rangle\langle n|\mathbf{F} ,
\end{equation}
that is $\{\mathbf{F}^\dag|n\rangle\}_{n=-s}^s$ is an eigenbasis for 
$\hat{\mathbf{p}}$.
\subsection{A $d$-dimensional version of Pauli operators}
The two-dimensional Pauli operator \ $Z:\mathbb{C}^2\!\longrightarrow\!\mathbb{C}^2,$
\begin{equation}
\begin{array}{l}
Z|0\rangle\!=\!|0\rangle,\\
Z|1\rangle\!=\!-|1\rangle,
\end{array},
\end{equation}
admits \cite{Vourdas04} the  general version \ $\mathbf{Z}\!:\!\mathcal{H}\!\longrightarrow\!\mathcal{H},$
\begin{equation}
\mathbf{Z}|n\rangle\!=\!\mathrm{e}^{\frac{2\pi \mathrm{i}}{d}n}|n\rangle 
\end{equation}
satisfying the relations
\begin{equation}
\begin{array}{l}
\mathbf{Z}\!=\!\mathbf{Z}\mathbb{I}\!=\!\mathbf{Z}\sum\limits_{n=-s}^s |n\rangle\langle n|\!=\!\sum\limits_{n=-s}^s \mathbf{Z}|n\rangle\langle n|\!=\!\sum\limits_{n=-s}^s \mathrm{e}^{\frac{2\pi \mathrm{i}}{d}n}|n\rangle\langle n|,\\[2mm]
(\mathbf{Z}\psi)(n)\!=\!\langle n|\mathbf{Z}|\psi\rangle
\!=\!\sum\limits_{m=-s}^s \mathrm{e}^{\frac{2\pi \mathrm{i}}{d}m}\langle n|m\rangle\langle m|\psi\rangle
\!=\!\mathrm{e}^{\frac{2\pi \mathrm{i}}{d}n}\psi(n),\\[2mm]
\mathrm{e}^{\frac{2\pi \mathrm{i}}{d}\hat{\mathbf{q}}} \!=\!
\mathrm{e}^{\frac{2\pi \mathrm{i}}{d}\hat{\mathbf{q}}}\mathbb{I} \!=\!
\mathrm{e}^{\frac{2\pi \mathrm{i}}{d}\hat{\mathbf{q}}}\sum\limits_{n=-s}^s |n\rangle\langle n| \!=\!
\sum\limits_{n=-s}^s \mathrm{e}^{\frac{2\pi \mathrm{i}}{d}\hat{\mathbf{q}}}|n\rangle\langle n| \!=\!
\sum\limits_{n=-s}^s \mathrm{e}^{\frac{2\pi \mathrm{i}}{d}n}|n\rangle\langle n|\!=\!\mathbf{Z}.
\end{array}
\end{equation}

The two-dimensional Pauli operator \ $X:\mathbb{C}^2\!\longrightarrow\!\mathbb{C}^2,$
\begin{equation}
\begin{array}{l}
X|0\rangle\!=\!|1\rangle,\\
X|1\rangle\!=\!|0\rangle,
\end{array}
\end{equation}
admits \cite{Vourdas04} the general version \ $\mathbf{X}\!:\!\mathcal{H}\!\longrightarrow\!\mathcal{H},$
\begin{equation}
\mathbf{X}|n\rangle\!=\!|n\!+\!1\rangle 
\end{equation}
where the addition is modulo $d$, and we choose
$n\!+\!1\!\in\!\{-\!s,\!-s\!+\!1,...,s\!-\!1,s\}$. \ It has the properties
\begin{equation}
\begin{array}{l}
\mathbf{X}\!=\!\mathbf{X}\mathbb{I}\!=\!\mathbf{X}\sum\limits_{n=-s}^s |n\rangle\langle n|\!=\!\sum\limits_{n=-s}^s \mathbf{X}|n\rangle\langle n|\!=\!\sum\limits_{n=-s}^s |n\!+\!1\rangle\langle n|,\\[2mm]
(\mathbf{X}\psi)(n)\!=\!\langle n|\mathbf{X}|\psi\rangle
\!=\!\sum\limits_{m=-s}^s \langle n|m\!+\!1\rangle\langle m|\psi\rangle
\!=\!\psi(n\!-\!1),\\[2mm]
\mathrm{e}^{-\frac{2\pi \mathrm{i}}{d}\hat{\mathbf{p}}} \!=\!
\mathbf{F}^\dag\mathrm{e}^{-\frac{2\pi \mathrm{i}}{d}\hat{\mathbf{q}}} \mathbf{F}\!=\!\mbox{\small $\frac{1}{\sqrt{d}}$}
\mathbf{F}^\dag\mathrm{e}^{-\frac{2\pi \mathrm{i}}{d}\hat{\mathbf{q}}} \sum\limits_{n,k=-s}^s \mathrm{e}^{-\frac{2\pi \mathrm{i}}{d}kn}\, |k\rangle\langle n|\!=\!
\mbox{\small $\frac{1}{\sqrt{d}}$}
\mathbf{F}^\dag \sum\limits_{n,k=-s}^s \mathrm{e}^{-\frac{2\pi \mathrm{i}}{d}kn}\,\mathrm{e}^{-\frac{2\pi \mathrm{i}}{d}k} |k\rangle\langle n|\\[1mm]
\qquad \ \ \  =\!\mbox{\small $\frac{1}{\sqrt{d}}$}
\sum\limits_{n,k=-s}^s \mathrm{e}^{-\frac{2\pi \mathrm{i}}{d}k(n+1)}\, \mathbf{F}^\dag |k\rangle\langle n|\!=\!
\mbox{\small $\frac{1}{d}$}
\sum\limits_{n,k,m=-s}^s \mathrm{e}^{-\frac{2\pi \mathrm{i}}{d}k(n+1)}\, \mathrm{e}^{\frac{2\pi \mathrm{i}}{d}km}\, |m\rangle\langle n|\\[1mm]
\qquad \ \ \ =\!\sum\limits_{n,m=-s}^s \mbox{\small $\frac{1}{d}$}
\sum\limits_{k=-s}^s \mathrm{e}^{\frac{2\pi \mathrm{i}}{d}k(m-n-1)}\,  |m\rangle\langle n|\!=\!
\sum\limits_{n,m=-s}^s \delta_{m,n+1}\,  |m\rangle\langle n|\!=\!
\sum\limits_{n=-s}^s  |n\!+\!1\rangle\langle n|\!=\!\mathbf{X}.
\end{array}
\end{equation}
The operators $\mathbf{Z},\ \mathbf{X}:\!\mathcal{H}\!\longrightarrow\!\mathcal{H}$ are unitary transformations:
\begin{equation}
\begin{array}{l}
 \langle \mathbf{Z}\varphi,\mathbf{Z}\psi\rangle\!
 =\!\sum\limits_{n=-s}^s\overline{\mathbf{Z}\varphi(n)}\  
 \mathbf{Z}\psi(n)\!=\!\sum\limits_{n=-s}^s\overline{\varphi(n)}\, 
 \psi(n)\!=\!\langle \varphi,\psi\rangle,\\[1mm]
 \langle \mathbf{X}\varphi,\mathbf{X}\psi\rangle\!
 =\!\sum\limits_{n=-s}^s\overline{\mathbf{X}\varphi(n)}\  
 \mathbf{X}\psi(n)\!=\!\sum\limits_{n=-s}^s\overline{\varphi(n\!-\!1)}\, 
 \psi(n\!-\!1)\!=\!\langle \varphi,\psi\rangle.
\end{array}
\end{equation}
\subsection{A discrete version of the displacement operators}
From
\begin{equation}
\begin{array}{l}
 \mathbf{Z}\mathbf{X}\! =\!\sum\limits_{n,m=-s}^s\mathrm{e}^{\frac{2\pi \mathrm{i}}{d}n}
 |n\rangle\langle n|m\!+\!1\rangle\langle m|
 \! =\!\sum\limits_{m=-s}^s\mathrm{e}^{\frac{2\pi \mathrm{i}}{d}(m+1)}  |m\!+\!1\rangle\langle m|,\\[1mm]
 \mathbf{X}\mathbf{Z}\! =\!\sum\limits_{n,m=-s}^s\mathrm{e}^{\frac{2\pi \mathrm{i}}{d}n}
 |m\!+\!1\rangle\langle m|n\rangle\langle n|
 \! =\!\sum\limits_{m=-s}^s\mathrm{e}^{\frac{2\pi \mathrm{i}}{d}m}  |m\!+\!1\rangle\langle m|
\end{array}
\end{equation}
it follows that
\begin{equation}
\mathbf{Z}\mathbf{X}\! =\!\mathrm{e}^{\frac{2\pi \mathrm{i}}{d}}\,
\mathbf{X}\mathbf{Z},
\end{equation}
and the relations
\begin{equation}
\begin{array}{c}
\mathrm{e}^{-\frac{\pi \mathrm{i}}{d}nk}\, \mathbf{Z}^k\mathbf{X}^n\! =\!\mathrm{e}^{\frac{\pi \mathrm{i}}{d}nk}\,
\mathbf{X}^n\mathbf{Z}^k,\\[1mm]
\mathrm{e}^{-\frac{\pi \mathrm{i}}{d}nk}\,\mathrm{e}^{\frac{2\pi \mathrm{i}}{d}k\hat{\mathbf{q}}}\,\mathrm{e}^{-\frac{2\pi \mathrm{i}}{d}n\hat{\mathbf{p}}} \!=\!\mathrm{e}^{\frac{\pi \mathrm{i}}{d}nk}\,\mathrm{e}^{-\frac{2\pi \mathrm{i}}{d}n\hat{\mathbf{p}}}\,\mathrm{e}^{\frac{2\pi \mathrm{i}}{d}k\hat{\mathbf{q}}}.
\end{array}
\end{equation}

In the continuous-variable case, the displacement operators are defined as
\begin{equation}
 \mathbf{D}(q,p)\!=\!\mathrm{e}^{-\frac{\mathrm{i}}{2\hbar}qp}\,\mathrm{e}^{\frac{\mathrm{i}}{\hbar}p\hat{q}}\,\mathrm{e}^{-\frac{\mathrm{i}}{\hbar}q\hat{p}} 
 \!=\!\mathrm{e}^{-\frac{\pi\mathrm{i}}{h}qp}\,\mathrm{e}^{\frac{2\pi\mathrm{i}}{h}p\hat{q}}\,\mathrm{e}^{-\frac{2\pi\mathrm{i}}{h}q\hat{p}}.
\end{equation}
By following the analogy with the continuous case, the discrete {\em displacement operators} are defined as  \ $D(n ,k )\!:\!\mathcal{H}\!\rightarrow\!\mathcal{H},$
\begin{equation}
 \mathbf{D}(n ,k )\!=\!\mathrm{e}^{-\frac{\pi \mathrm{i}}{d}nk}\,\mathbf{Z}^k\,\mathbf{X}^n\! =\!\mathrm{e}^{-\frac{\pi \mathrm{i}}{d}nk}\,\mathrm{e}^{\frac{2\pi \mathrm{i}}{d}k\hat{\mathbf{q}}}\,\mathrm{e}^{-\frac{2\pi \mathrm{i}}{d}n\hat{\mathbf{p}}}. 
\end{equation}
They satisfy the relations \cite{Vourdas04}
\begin{equation}
\begin{array}{l}
 \mathbf{D}(n ,k )\psi (m)=\mathrm{e}^{-\frac{\pi \mathrm{i}}{d}nk}\, \mathrm{e}^{\frac{2\pi \mathrm{i}}{d}km}\, \psi (m\!-\!n ),\\[2mm]
 \mathbf{D}(n ,k )\, \mathbf{D}(m ,\ell )\!=\!\mathrm{e}^{\frac{\pi \mathrm{i}}
{d}(km-n\ell)} \ \mathbf{D}(n\!+\!m ,k\!+\!\ell )  \\[2mm]
\mathbf{D}(n ,k )\, \mathbf{D}(m ,\ell )\!=\!\mathrm{e}^{\frac{2\pi \mathrm{i}} 
{d}(km-n\ell)}\ \mathbf{D}(m ,\ell )\ \mathbf{D}(n ,k )\\[2mm]
\mathbf{D}^\dag(n ,k )\!=\!\mathbf{D}(-n ,-k )\!=\!\mathbf{D}^{-1}(n ,k ).
\end{array}
\end{equation}

%
%
%
\section{Discrete version of the position-momentum commutation relation}
 \begin{table}[h]
\caption{\label{Eigenval}
Eigenvalues of $\mathbf{C}\!=\![\hat{\mathbf{q}},\hat{\mathbf{p}}]-{\rm i}\frac{d}{2\pi }$ in two cases.
}
\begin{tabular}{cc}
\hline
Case $d=11$   & \qquad Case $d=31$   \\
\hline
$\begin{array}{r}
-1.48548\!\times\! 10^{-8}\,{\rm i}\\
7.96337\!\times\! 10^{-7}\,{\rm i}\\
2.04883\!\times\!10^{-5}\,{\rm i}\\
3.36098\!\times\!10^{-4}\,{\rm i}\\
3.93615\!\times\!10^{-3}\,{\rm i}\\
3.49009\!\times\! 10^{-2}\,{\rm i}\\
-0.240939\,{\rm i}\\
1.33714\,{\rm i}\\
-5.45524\,{\rm i}\\
19.9934\,{\rm i}\\
-34.9234\,{\rm i}
\end{array}$     &
\quad $\begin{array}{r}
-6.8034\!\times\!10^{-16}\,{\rm i}\\
-1.28585\!\times\!10^{-15}\,{\rm i}\\
2.07899\!\times\!10^{-15}\,{\rm i}\\
-3.04408\!\times\!10^{-15}\,{\rm i}\\
3.57322\!\times\!10^{-15}\,{\rm i}\\
5.70417\!\times\!10^{-15}\,{\rm i}\\
-1.12090\!\times\!10^{-14}\,{\rm i}\\
1.58666\!\times\!10^{-14}\,{\rm i}\\
-2.13350\!\times\!10^{-14}\,{\rm i}\\
1.13714\!\times\!10^{-13}\,{\rm i}\\
-1.41394\!\times\!10^{-12}\,{\rm i}\\
1.73886\!\times\!10^{-11}\,{\rm i}\\
-1.91428\!\times\!10^{-10}\,{\rm i}\\
1.9093\!\times\!10^{-9}\,{\rm i}\\
-1.73552\!\times\!10^{-8}\,{\rm i}\\
\end{array}$\quad   
\quad $\begin{array}{r}
1.44395\!\times\!10^{-7}\,{\rm i}\\
-1.10354\!\times\!10^{-6}\,{\rm i}\\
7.76978\!\times\!10^{-6}\,{\rm i}\\
-0.000050515\,{\rm i}\\
0.000303805\,{\rm i}\\
-0.00169225\,{\rm i}\\
0.00873614\,{\rm i}\\
-0.0417956\,{\rm i}\\
0.185225\,{\rm i}\\
-0.757941\,{\rm i}\\
2.86832\,{\rm i}\\
-9.80352\,{\rm i}\\
31.5302\,{\rm i}\\
-80.4393\,{\rm i}\\
214.181\,{\rm i}\\
-310.677\,{\rm i}\\
\end{array}$\\
\hline
\end{tabular}
\end{table}
%
%
In the continuous case, the usual position and momentum satisfy 
the remarkable relation \ $[\hat{q},\hat{p}]={\rm i}\hbar$, that is
\begin{equation}\label{qp-pq}
[\hat{q},\hat{p}]={\rm i}\mbox{\small $\frac{h}{2\pi }$}.
\end{equation}
This means that
\begin{equation}
\hat{q}(\hat{p}\Psi)-\hat{p}(\hat{q}\Psi)\!=\!{\rm i}\mbox{\small $\frac{h}{2\pi }$}\Psi
\end{equation}
for any $\Psi$ from the subspace
\begin{equation}
\left\{\, \Psi \!\in \! D_q\cap D_p\ \ |\ \ \hat{q}\Psi \!\in\!D_p\ \ \mbox{and}\ \ 
\hat{p}\Psi \!\in\!D_q \, \right\}.
\end{equation}

If we compare the continuous and discrete version of the Fourier transform, then we arrive at the conclusion that, in the discrete case, the role of Planck constant is played by the dimension $d$ of the Hilbert space. Consequently, the discrete version of the relation (\ref{qp-pq}) is
\begin{equation}\label{qp-pqd}
[\hat{\mathbf{q}},\hat{\mathbf{p}}]={\rm i}\mbox{\small $\frac{d}{2\pi }$}
\end{equation}
One can see that this relation is not satisfied in the whole $\mathcal{H}$, but
by computing the eigenvalues of the operator \ $\mathbf{C}\!:\!\mathcal{H}\!\rightarrow\!\mathcal{H},$
\begin{equation}
\begin{array}{l}
\mathbf{C}\!=\![\hat{\mathbf{q}},\hat{\mathbf{p}}]-{\rm i}\frac{d}{2\pi }
\end{array}
\end{equation}
with the matrix elements
\begin{equation}
\begin{array}{l}
\langle n|\mathbf{C}|m\rangle\!=\!\frac{1}{d}\sum\limits_{k=-s}^s k(n\!-\!m)\, \mathrm{e}^{\frac{2\pi \mathrm{i}}{d}k(n-m) }-{\rm i}\mbox{\small $\frac{d}{2\pi }$}\delta_{nm}
\end{array}
\end{equation}
for different values of $d$, one arrives at the conclusion 
that the eigenvalues $\lambda$  with $|\lambda|<0.001$
represent $36\%$ in the case $d\!=\!11$, represent $64\%$ in the case $d\!=\!31$, represent $77\%$ in the case $d\!=\!61$ and $84\%$ in the case 
$d\!=\!101$ (see Table I). 
So, for $d$ large enough, a significant part of the eigenvalues of 
$[\hat{\mathbf{q}},\hat{\mathbf{p}}]-{\rm i}\frac{d}{2\pi }$
are almost null.

For any $\varepsilon\!>\!0$, there exists $d_\varepsilon $ such that the set
\begin{equation}
\mathcal{S}_\varepsilon\!=\!\left\{\, \psi\!\in\!\mathcal{H}\ |\ ||\psi||\!=\!1\ \mbox{and}\ ||\mathbf{C}\psi||\!<\!\varepsilon \ \right\}
\end{equation}
is non-empty for $d\!=\!\mathrm{dim}(\mathcal{H})\!>\!d_\varepsilon$.
If $\varepsilon\!>\!0$ is small enough, then we can consider that
\begin{equation}\label{qppq}
\begin{array}{l}
[\hat{\mathbf{q}},\hat{\mathbf{p}}]\psi \approx{\rm i}\frac{d}{2\pi }\psi \quad \mbox{for}\ \ \psi\!\in\!\mathcal{S}_\varepsilon.
\end{array}
\end{equation}
If the states $\psi_1$, $\psi_2$, ... , $\psi_k$ from $\mathcal{S}_\varepsilon$ form an orthonormal system  
and $\psi\!=\!\sum\limits_{m=1}^k\alpha _m \psi_m$ is such that $||\psi||\!=\!1$, then
\begin{equation}
||\mathbf{C}\psi||\!=\!\left|\left|\sum\limits_{m=1}^k\alpha _m \mathbf{C}\psi_m \right|\right|\!\leq\!\sum\limits_{m=1}^k |\alpha _m|\, \parallel \mathbf{C}\psi_m\parallel\!
\leq\!\varepsilon\sqrt{\sum\limits_{m=1}^k |\alpha _m|^2}\sqrt{\sum\limits_{m=1}^k 1^2}
\!\leq \!\varepsilon\sqrt{k},
\end{equation}
that is $\psi\!\in\!\mathcal{S}_{\varepsilon\sqrt{k}}$.
\subsection{A spectral decomposition of the Fourier transform}
By using the relation
\[
(\mathbf{F}\hat{\mathbf{q}}\mathbf{F}^\dag)\psi (n)\!=\!
\frac{1}{d}\sum\limits_{m,k=-s}^s\mathrm{e}^{\frac{2\pi \mathrm{i}}{d}m(k-n)} \, m\, \psi(k)\!
=\!\frac{1}{d}\sum\limits_{m,k=-s}^s\mathrm{e}^{-\frac{2\pi \mathrm{i}}{d}m(k-n)} \, (-m)\, \psi(k)\!
=\!-(\mathbf{F}^\dag\hat{\mathbf{q}}\mathbf{F})\psi (n)\!=\!-\hat{\mathbf{p}}\psi (n)
\]
we get \ $\mathbf{F}\hat{\mathbf{q}}\mathbf{F}^\dag\!=\!-\hat{\mathbf{p}}$ \ and
\[  \mathbf{F}[\hat{\mathbf{q}},\hat{\mathbf{p}}]\!
  =\!\mathbf{F} \hat{\mathbf{q}}\,\hat{\mathbf{p}}-
  \mathbf{F} \hat{\mathbf{p}}\,\hat{\mathbf{q}}\!=\!
\mathbf{F} \hat{\mathbf{q}}\,\mathbf{F}^\dag\hat{\mathbf{F}}-
 \mathbf{F} \mathbf{F}^\dag\hat{\mathbf{q}}\mathbf{F}\,\hat{\mathbf{q}}
  \!=\!
 -\hat{\mathbf{p}}\,\hat{\mathbf{q}}\mathbf{F}-
  \hat{\mathbf{q}}\mathbf{F}\,\hat{\mathbf{q}}\mathbf{F}^\dag \mathbf{F}\!=\![\hat{\mathbf{q}},\hat{\mathbf{p}}]\mathbf{F},
\]
that is
\begin{equation}\label{fqppq}
\mathbf{F}[\hat{\mathbf{q}},\hat{\mathbf{p}}]\!=\!
[\hat{\mathbf{q}},\hat{\mathbf{p}}]\mathbf{F}.
\end{equation}
From the relation  
\begin{equation}
\begin{array}{l}
||\mathbf{C}\mathbf{F}\psi ||\!=\!||\mathbf{F}\mathbf{C}\psi ||\!=\!||\mathbf{C}\psi ||,
\end{array}
\end{equation}
it follows that any set $\mathcal{S}_\varepsilon$ is Fourier invariant, that is
\begin{equation}
\psi\!\in\!\mathcal{S}_\varepsilon\quad \Rightarrow \quad \mathbf{F}\psi\!\in\!\mathcal{S}_\varepsilon.
\end{equation}

Let $\lambda _0$, $\lambda _1$, ..., $\lambda _{d-1}$ be the  eigenvalues of $\mathbf{C}\!=\![\hat{\mathbf{q}},\hat{\mathbf{p}}]-{\rm i}\frac{d}{2\pi }$, considered in the increasing order of their modulus ($|\lambda _0|\leq |\lambda _1|\leq ...\leq |\lambda _{d-1}|$), and let
$\varphi_0$, $\varphi_1$, ... , $\varphi_{d-1}$  be the corresponding eigenfunctions, that is
\begin{equation}
\mathbf{C}\varphi_k\!=\!\lambda_k\ \varphi_\kappa.
\end{equation}
The relation 
\[
\mathbf{C}\varphi_k\!=\!\lambda_k\, \varphi_\kappa\quad 
\Rightarrow \quad
\mathbf{C}\mathbf{F}\varphi_k\!=\!\mathbf{F}\mathbf{C}\varphi_k\!=\!\lambda_k\ \mathbf{F}\,\varphi_\kappa
\]
%
 \begin{table}[h]
\caption{ \label{Fourier}
Case $d\!=\!11$: \ $\varphi_1,\, \varphi_2,...,\varphi_d$ are eigenstates of $\mathbf{F}$.
}
\begin{tabular}{rrrrr}
\hline
$k$ & \quad $\parallel \mathbf{F}\varphi_k\!-\!\varphi_k\parallel $ & \qquad $\parallel \mathbf{F}\varphi_k\!+\!\varphi_k\parallel $ & \qquad $\parallel \mathbf{F}\varphi_k\!-\!\mathrm{i} \varphi_k\parallel $ & \qquad $\parallel \mathbf{F}\varphi_k\!+\!\mathrm{i}\varphi_k\parallel $  \\
\hline
1 & $2.0\!\times\!10^{-10}$ & 2 & 1.4142 & 1.4142 \\
2 & 1.4142 & 1.4142 & 2 & $1.2\!\times\!10^{-10}$ \\
3 &  2 & $6.3\!\times\!10^{-11}$  & 1.4142  & 1.4142 \\
4 & 1.4142 &  1.4142  & $1.3\!\times\!10^{-12}$ & 2\\
5 &  $2.7\!\times\!10^{-13}$ & 2  & 1.4142 & 1.4142 \\
6 &  1.4142 & 1.4142 & 2 & $2.3\!\times\!10^{-14}$ \\
7 & 2 &  $1.3\!\times\!10^{-14}$ &   1.4142   &  1.4142  \\
8 & 1.4142 & 1.4142 & $3.4\!\times\!10^{-15}$ & 2\\
9 & $1.0\!\times\!10^{-15}$ & 2 & 1.4142   &  1.4142   \\
10 & 1.4142 & 1.4142   & 2 & $8.3\!\times\!10^{-16}$ \\
11 & 2 & $8.4\!\times\!10^{-16}$ & 1.4142 & 1.4142 \\
\hline
 \end{tabular}
\end{table}
%
and the numerical data (see Table \ref{Fourier}) suggest that
the states $\varphi_0$, $\varphi_1$, ... , $\varphi_{d-1}$ are also eigenstates of $\mathbf{F}$, namely \cite{Mehta87,Barker}
\begin{equation} \label{Fourierphi}
\mathbf{F}\varphi_n\!=\!(-\mathrm{i})^n\ \varphi_n.
\end{equation}
Particularly, the discrete Fourier transform admits the spectral decomposition
\begin{equation} 
\mathbf{F}\!=\!\sum_{n=0}^{d-1}(-\mathrm{i})^n\ |\varphi_n\rangle\langle\varphi_n|.
\end{equation}
For any $\varepsilon >0$ and $d$ high enough, some of the states
$\varphi_0$, $\varphi_1$, ... , $\varphi_{d-1}$ belong to  $\mathcal{S}_\varepsilon$. For example (see Table \ref{Eigenval}), in the case $d\!=\!31$, the states
$\varphi_0$, $\varphi_1$, ... , $\varphi_{19}$ belong to $\mathcal{S}_{0.001}$.
\subsection{A discrete version of the uncertainty relation}
A direct consequence of (\ref{qppq}) is
\begin{equation}
|\langle \psi |[\hat{\mathbf{q}},\hat{\mathbf{p}}]|\psi\rangle |\approx \mbox{\small $\frac{d}{2\pi }$} \quad \mbox{for}\ \ \psi\!\in\!\mathcal{S}_\varepsilon.
\end{equation}
From the Robertson-Schr\"{o}dinger 
uncertainty relation, it follows that the inequality
\begin{equation}
\Delta\hat{\mathbf{q}}\ \Delta\hat{\mathbf{p}}\geq \frac{|\langle \psi |[\hat{\mathbf{q}},\hat{\mathbf{p}}]|\psi |}{2} \quad \mbox{for}\ \ \psi\!\in\!\mathcal{S}_\varepsilon,
\end{equation}
that is
\begin{equation}
\Delta\hat{\mathbf{q}}\ \Delta\hat{\mathbf{p}}\geq \mbox{\small $\frac{d}{4\pi }$}
\end{equation}
is satisfied with approximation in $\mathcal{S}_\varepsilon$.
See the particular cases presented in the Tables V-XIII and Figure \ref{modulg}.
\subsection{Commutation relation of discrete creation-annihilation operators}
The operators 
\begin{equation}
\begin{array}{r}
\hat{\mathbf{a}}\!=\!\sqrt{\frac{\pi}{d}}(\hat{\mathbf{q}}\!+\!{\rm i}\hat{\mathbf{p}}),\\[2mm]
\hat{\mathbf{a}}^\dag\!=\!\sqrt{\frac{\pi}{d}}(\hat{\mathbf{q}}\!-\!{\rm i}\hat{\mathbf{p}}),
\end{array}
\end{equation}
can be regarded as a discrete version of the {\em creation} (also called  {\rm raising}) and annihilation (also called  {\rm lowering})  operators
\begin{equation}
\begin{array}{r}
\hat{a}\!=\!\sqrt{\frac{\pi}{h}}(\hat{q}\!+\!{\rm i}\hat{p}),\\[2mm]
\hat{a}^\dag\!=\!\sqrt{\frac{\pi}{h}}(\hat{q}\!-\!{\rm i}\hat{p}),
\end{array}
\end{equation}
used in the continuous variable case.
For small $\varepsilon\!>\!0$ and any $\psi\!\in\!\mathcal{S}_\varepsilon$, we have 
$[\hat{\mathbf{q}},\hat{\mathbf{p}}]\psi\approx \mathrm{i}\frac{d}{2\pi}\psi\ \Rightarrow \ [\mathbf{a}^\dag,\mathbf{a}]\psi \approx \psi$, that is
\begin{equation}
[\hat{\mathbf{a}},\hat{\mathbf{a}}^\dag]\!\approx\! 1\qquad \mbox{in } \ \mathcal{S}_\varepsilon .
\end{equation}

In the continuous-variable case, by using Baker–Campbell–Hausdorff formula, one gets
\begin{equation}
[\hat{q},\hat{p}]={\rm i}\mbox{\small $\frac{h}{2\pi }$}  \quad  \Longrightarrow  \quad  D(q,p)
 \!=\!\mathrm{e}^{-\frac{\pi\mathrm{i}}{h}qp}\,\mathrm{e}^{\frac{2\pi\mathrm{i}}{h}p\hat{q}}\,\mathrm{e}^{-\frac{2\pi\mathrm{i}}{h}q\hat{p}}\!=\!\mathrm{e}^{\frac{2\pi\mathrm{i}}{h}(p\hat{q}-q\hat{p})}.
\end{equation}
By denoting $\alpha\!=\!\sqrt{\frac{\pi}{h}}(q\!+\!\mathrm{i}p),$
this relation can be written as
\begin{equation}
D(\alpha) \!=\!\mathrm{e}^{\alpha\hat{a}^\dag-\bar \alpha\hat{a}},
\end{equation}
It is an open issue if 
\begin{equation}
[\hat{\mathbf{q}},\hat{\mathbf{p}}]\psi\approx \mathrm{i}\frac{d}{2\pi}\psi \quad \Longrightarrow  \quad \begin{array}{l}\mathbf{D}(n,k)\psi
 \!\approx\!\mathrm{e}^{\frac{2\pi\mathrm{i}}{d}(k\hat{\mathbf{q}}-n\hat{\mathbf{p}})}\psi,\\[1mm]
\mathbf{D}(\alpha)\psi \!\approx\!\mathrm{e}^{\alpha\hat{\mathbf{a}}^\dag-\bar \alpha\hat{\mathbf{a}}}\psi 
 \qquad \qquad \qquad \mbox{for } \ \alpha\!=\!\sqrt{\frac{\pi}{d}}(n\!+\!\mathrm{i}k).
 \end{array}
\end{equation}

%
%
%
%
\section{Some remarkable elements of $\mathcal{S}_\varepsilon$ and their properties}
\subsection{Discrete-variable Gaussian functions}
%
%
\begin{figure}[h]
\includegraphics[scale=1.5]{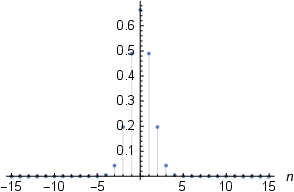}\qquad
\includegraphics[scale=1.5]{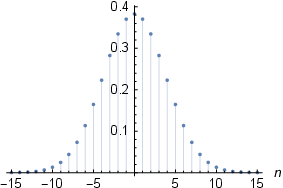}
\caption{Gaussian functions  $\mathbf{g}_3$ and $\mathbf{g}_{1/3} $  in the case $d\!=\!31$. }
\end{figure}
%

For any $\kappa \!\in\!(0,\infty)$, the Gaussian function  
$g_\kappa :\mathbb{R}\rightarrow \mathbb{R}$, 
\begin{equation}
\begin{array}{l}
g_\kappa (q)\!=\! \mathrm{e}^{-\frac{\kappa \pi}{h}q^2},
\end{array}
\end{equation}
satisfies the relation 
\begin{equation}\label{Fgkc}
F[g_\kappa]\!=\!\mbox{\small $\frac{1}{\sqrt{\kappa}}$}\, g_{1/\kappa}.
\end{equation}
The corresponding {\it discrete-variable Gaussian function} \ 
$\mathfrak{g}_\kappa \!:\!\{-s,\!-s\!+\!\!1,...,s\!-\!\!1,\!s\}\!\rightarrow \!\mathbb{R},$
\begin{equation}\label{gkappa}
\begin{array}{l}
 \mathfrak{g}_\kappa(n)=\sum\limits_{\alpha=-\infty}^\infty\mathrm{e}^{-\frac{\kappa \pi}{d}(n+\alpha d)^2},
\end{array}
\end{equation}
satisfies the relation \cite{Vourdas04,Cotfas12}
\begin{equation}
\mathbf{F}[\mathfrak{g}_\kappa]\!=\!\mbox{\small $\frac{1}{\sqrt{\kappa}}$}\ \mathfrak{g}_{1/\kappa}
\end{equation}
similar to (\ref{Fgkc}). It is a particular case of Jacobi $\theta$ function \cite{Vourdas04,Mehta87}. Because $\mathbf{F}$ is a unitary transform, the corresponding normalized discrete Gaussian functions (see Figure 1)
\begin{equation}
\mathbf{g}_\kappa\!=\!\frac{\mathfrak{g}_\kappa}{||\mathfrak{g}_\kappa||}\quad\mbox{where} \ \ 
||\mathfrak{g}_\kappa||\!=\!\sqrt{\sum\limits_{n=-s}^s (\mathfrak{g}_\kappa(n))^2},
\end{equation}
satisfy the relation
\begin{equation} \label{Fgkappa}
\mathbf{F}[\mathbf{g}_\kappa]\!=\!\mathbf{g}_{1/\kappa}.
\end{equation}

\begin{figure}[h]
\includegraphics[scale=1.2]{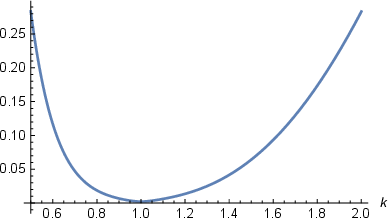}\qquad
\includegraphics[scale=1.2]{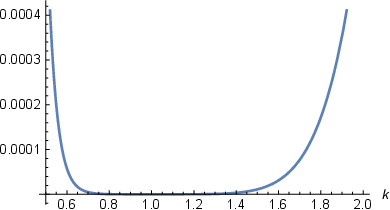}
\caption{\label{normg} Norm \ $\parallel ([\hat{\mathbf{q}},\hat{\mathbf{p}}]-{\rm i}\frac{d}{2\pi })\mathbf{g}_k||$  for $d\!=\!11$ and $d\!=\!31$. }
\end{figure}

\begin{figure}[h]
\includegraphics[scale=1.2]{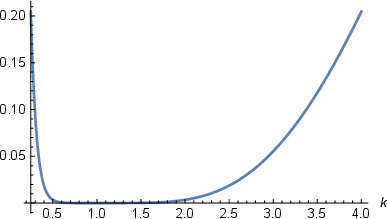}\qquad
\includegraphics[scale=1.2]{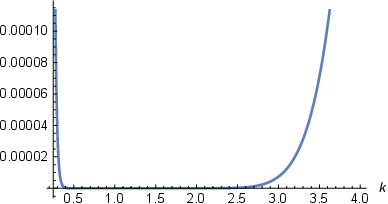}
\caption{\label{modulg} Modulus   $|\, |\langle \mathbf{g}_k |[\hat{\mathbf{q}},\hat{\mathbf{p}}]|\mathbf{g}_k\rangle |-\frac{d}{2\pi }\ |$  for  $d\!=\!11$ and $d\!=\!31$. }
\end{figure}

For $d$ large enough and small $\delta >0$, there exists 
$\eta >0$ such that, for $\kappa \!\in\!(\frac{1}{\eta},\eta)$, the most 
significant coordinates of $\mathbf{g}_\kappa$ in the eigenbasis 
$\{\varphi_1,\, \varphi_2,\, ...,\, \varphi_d\}$ of $\mathbf{C}\!=\![\hat{\mathbf{q}},\hat{\mathbf{p}}]-{\rm i}\frac{d}{2\pi }$ are those corresponding to the functions $\varphi_k$ with $|\lambda_k|<\delta$ (see Table \ref{coordGauss}). We can consider that 
\begin{equation}
[\hat{\mathbf{q}},\hat{\mathbf{p}}]\approx{\rm i}\mbox{\small $\frac{d}{2\pi }$}\quad \mbox{for}\ \mathbf{g}_\kappa 
 \ \mbox{with} \ \ \kappa \!\in\!(\mbox{\small $\frac{1}{\eta}$},\eta)
\end{equation}
(see Tables \ref{coordGauss}, \ \ref{normGauss} and Figure \ref{normg}). From (\ref{Fourierphi}), we get
\begin{equation}\label{phink}
|\langle \varphi_n|\mathbf{g}_\kappa\rangle |\!=\!
|\langle \theta_n \varphi_n|\mathbf{g}_\kappa\rangle |\!=\!
|\langle \mathbf{F} \varphi_n|\mathbf{g}_\kappa\rangle |\!=\!
|\langle  \varphi_n|\mathbf{g}_{1/\kappa}\rangle |,
\end{equation}
in agreement with the numerical data from Table \ref{coordGauss}.
 \begin{table*}[h]
\caption{\label{coordGauss}
Case $d\!=\!11$: for certain $\kappa$, the Gaussian $\mathbf{g}_\kappa$ is mainly a linear combination of 
$\varphi_1,$  $\varphi_2,\,\varphi_3,\,\varphi_4,\,\varphi_5$.
}
\begin{tabular}{rrrrrr}
\hline
 & \qquad $\kappa\!=\!1$ \qquad \qquad& \qquad $\kappa\!=\!2$ \qquad \qquad& \qquad $\kappa\!=\!1/2$ \qquad \qquad& \qquad $\kappa\!=\!3$ \qquad \qquad &  \quad $\kappa\!=\!1/3$ \quad \mbox{} \\
\hline
$|\langle \varphi_0|\mathbf{g}_\kappa\rangle |$  &  0.999968 \qquad & 0.970268 \qquad & 0.970268 \qquad & 0.926811 \qquad & 0.926811\\
$|\langle \varphi_1|\mathbf{g}_\kappa\rangle |$ & $8.79611\!\times\!10^{-11}$  & $8.56453\!\times\!10^{-11}$   & $8.50632\!\times\!10^{-11}$  & $8.19559\!\times\!10^{-11}$  & $8.11208\!\times\!10^{-8}$ \\
$|\langle \varphi_2|\mathbf{g}_\kappa\rangle |$ & $3.14918\!\times\!10^{-11}$  & 0.228139 & 0.228139 & 0.327299 & 0.327299 \\
$|\langle \varphi_3|\mathbf{g}_\kappa\rangle |$ & $3.83482\!\times\!10^{-13}$  & $3.8548\!\times\!10^{-13}$ & $3.58495\!\times\!10^{-13}$ & $3.74436\!\times\!10^{-13}$ & $3.36051\!\times\!10^{-13}$ \\
$|\langle \varphi_4|\mathbf{g}_\kappa\rangle |$ & $0.00798631$   & 0.0736351 & 0.0736351 & 0.151392 & 0.151392 \\
$|\langle \varphi_5|\mathbf{g}_\kappa\rangle |$ & $8.61761\!\times\!10^{-15}$   & $9.0834\!\times\!10^{-15}$  & $8.25462\!\times\!10^{-15}$  & $9.15919\!\times\!10^{-15}$ & $8.20907\!\times\!10^{-15}$  \\
$|\langle \varphi_6|\mathbf{g}_\kappa\rangle |$ & $2.78121\!\times\!10^{-15}$   & 0.0288995 & 0.0288995 & 0.0830314 & 0.0830314 \\
$|\langle \varphi_7|\mathbf{g}_\kappa\rangle |$ & $1.82021\!\times\!10^{-16}$   & $1.7344\!\times\!10^{-16}$  & $2.47453\!\times\!10^{-16}$ & $1.32587\!\times\!10^{-16}$ & $3.49103\!\times\!10^{-16}$\\
$|\langle \varphi_8|\mathbf{g}_\kappa\rangle |$ & $0.000416018$   & 0.0146752 & 0.0146752 & 0.0549712 & 0.0549712 \\
$|\langle \varphi_{9}|\mathbf{g}_\kappa\rangle |$ & $5.0058\!\times\!10^{-17}$   & $4.89024\!\times\!10^{-17}$ & $6.26338\!\times\!10^{-17}$ & $4.68507\!\times\!10^{-17}$ & $7.58437\!\times\!10^{-17}$\\
$|\langle \varphi_{10}|\mathbf{g}_\kappa\rangle |$ & $1.98563\!\times\!10^{-16}$   & 0.00777324 & 0.00777324 & 0.0325688 & 0.0325688\\
\hline
\end{tabular}
\end{table*}
%
Since $\mathbf{F}$ is a unitary transform, we get 
\begin{equation}
||[\hat{\mathbf{q}},\hat{\mathbf{p}}]\psi -{\rm i}\frac{d}{2\pi }\psi||=||\mathbf{F}[\hat{\mathbf{q}},\hat{\mathbf{p}}]\psi -{\rm i}\frac{d}{2\pi }\mathbf{F}\psi||
=||[\hat{\mathbf{q}},\hat{\mathbf{p}}]\mathbf{F}\psi -{\rm i}\frac{d}{2\pi }\mathbf{F}\psi||,
\end{equation}
for any $\psi\!\in\!\mathcal{H}$. Particularly, we have
\begin{equation}\label{qpigk}
\parallel\mathbf{C}\mathbf{g}_\kappa\parallel
=\parallel\mathbf{C}\mathbf{g}_{1/\kappa}\parallel,
\end{equation}
in agreement with the numerical data from Table \ref{normGauss}.\\
 By using  (\ref{fqppq}) and (\ref{Fgkappa}), we get
\begin{equation}\label{g1kappa}
\langle \mathbf{g}_{1/\kappa} |[\hat{\mathbf{q}},\hat{\mathbf{p}}]|\mathbf{g}_{1/\kappa}\rangle \!=\!\langle \mathbf{F}\mathbf{g}_\kappa |\mathbf{F}[\hat{\mathbf{q}},\hat{\mathbf{p}}]|\mathbf{g}_\kappa\rangle \!=\!\langle \mathbf{g}_\kappa |[\hat{\mathbf{q}},\hat{\mathbf{p}}]|\mathbf{g}_\kappa\rangle ,
\end{equation}
in agreement with the numerical data from Table \ref{modulGauss}.
 \begin{table}[h]
\caption{\label{normGauss}  
$[\hat{\mathbf{q}},\hat{\mathbf{p}}]\mathbf{g}_\kappa\approx{\rm i}\mbox{\small $\frac{d}{2\pi }$}\mathbf{g}_\kappa$ is satisfied for certain $\mathbf{g}_\kappa$.
}
\begin{tabular}{crr}
\hline
$\kappa$ & & $||\left([\hat{\mathbf{q}},\hat{\mathbf{p}}]-{\rm i}\frac{d}{2\pi }\right)\mathbf{g}_\kappa||$\qquad\qquad\qquad\qquad\mbox{}  \\
& Case $d=11$   &  Case $d=31$   \\
\hline
1 & 0.0022697 &  $2.68152\!\times\!10^{-9}$\\
2 & 0.283112 & 0.000670943\\
1/2 & 0.283112 & 0.000670943\\
 3    &     & 0.0383181\\
 1/3   &   &  0.0383181\\
\hline
 \end{tabular}
\end{table}
 \begin{table} [h]
\caption{\label{modulGauss}
$|\langle \mathbf{g}_\kappa |[\hat{\mathbf{q}},\hat{\mathbf{p}}]|\mathbf{g}_\kappa\rangle | \approx \frac{d}{2\pi }$  is satisfied for certain $\mathbf{g}_\kappa$.
}
{\scriptsize
\begin{tabular}{crr}
\hline
$\kappa$ &  &$|\ |\langle \mathbf{g}_\kappa |[\hat{\mathbf{q}},\hat{\mathbf{p}}]|\mathbf{g}_\kappa\rangle |-\frac{d}{2\pi }\ |$\qquad\qquad\qquad\qquad\mbox{}  \\
& Case $d=11$   &  Case $d=31$   \\
\hline
1 & $1.21005\!\times\!10^{-6}$ &  $8.88178\!\times\!10^{-16}$\\
2 & 0.00350868 & $2.47852\!\times\!10^{-9}$\\
1/2 & 0.00350868 & $2.47852\!\times\!10^{-9}$\\
3   &  0.0552824  & $7.39351\!\times\!10^{-6}$\\
1/3   &  0.0552824  & $7.39351\!\times\!10^{-6}$\\
4  & 0.204735  & 0.000383435\\
1/4  & 0.204735  & 0.000383435\\
5  &    &  0.00397916\\
1/5  &    &  0.00397916\\
6  &    & 0.0185731\\
1/6  &    & 0.0185731\\
7  &   & 0.0550503\\
1/7  &   & 0.0550503\\
8  &    &  0.122973\\
1/8  &    &  0.122973\\
\hline
\end{tabular}
}
\end{table}

In conclusion, if the dimension $d$ of the Hilbert space describing
the quantum system (qudit) is large enough, there exists a family $\mathcal{G}_\delta\!=\!\left\{ \, \mathbf{g}_\kappa \ |\ \kappa \!\in\!(\mbox{\small $\frac{1}{\eta}$},\eta)\ \right\}$ of discrete-variable Gaussian states $\mathbf{g}_\kappa$ depending on a continuous parameter 
$\kappa \!\in\!(\mbox{\small $\frac{1}{\eta}$},\eta)$ such that the relations
\begin{equation}\label{2rel}
[\hat{\mathbf{q}},\hat{\mathbf{p}}]={\rm i}\mbox{\small $\frac{d}{2\pi }$}\quad \mbox{and}\quad \Delta\hat{\mathbf{q}}\ \Delta\hat{\mathbf{p}}\geq \mbox{\small $\frac{d}{4\pi }$}
\end{equation}
are satisfied with approximation for any $\mathbf{g}_\kappa\!\in\!\mathcal{G}_\delta$ (See Figure 2 and Figure 3). 
%
Any desired precision can be obtained by increasing the dimension $d$ of the Hilbert space and use of an adequate interval $(\mbox{\small $\frac{1}{\eta}$},\eta)$. In addition, the family $ \mathcal{G}_\delta$ is Fourier invariant: 
\begin{equation}
\mathbf{g}_\kappa\!\in\!\mathcal{G}_\delta\quad \Rightarrow \quad \mathbf{F}[\mathbf{g}_\kappa]\!=\!\mathbf{g}_{1/\kappa}\!\in\!\mathcal{G}_\delta .
\end{equation}
\subsection{Discrete version of Fock states}
It is known that the Hamiltonian of the harmonic oscillator can be written as
\begin{equation}
\hat{H}\!=\!\hat{a}^\dag \hat{a}\!+\!\frac{1}{2},
\end{equation}
and its eigenfunctions are the Fock states
\begin{equation}
\Psi_n\!=\!\frac{1}{\sqrt{n!}}(\hat{a}^\dag)^n\Psi_0
\end{equation}
where $\Psi_0(q)\!=\!\frac{1}{\sqrt[4]{\pi}}\mathrm{e}^{-\frac{1}{2}q^2}$ is the vacuum state.
By following the analogy with the continuous case, we define
\begin{equation}
\psi_{C,n}\!=\!\frac{1}{|| (\hat{\mathbf{a}}^\dag)^n\mathbf{g}||}(\hat{\mathbf{a}}^\dag)^n\mathbf{g},
\end{equation}
where $\mathbf{g}\!\equiv\!\mathbf{g}_1$ is the discrete version of the vacuum state.

By direct computation, we can remark that, for small $\varepsilon\!>\!0$ and 
$d$ high enough, some of the states $\psi_{C,n}$ belong to $\mathcal{S}_\varepsilon$. For example (see Table \ref{creation}), in the case $d\!=\!31$, the states 
$\psi_{C,0}$, $\psi_{C,1}$, ... , $\psi_{C,8}$ belong to $\mathcal{S}_{0.001}$.
 \begin{table}[h]
\caption{\label{creation} In the case $d\!=\!31$, the relations $[\hat{\mathbf{q}},\hat{\mathbf{p}}]\psi \approx{\rm i}\mbox{\small $\frac{d}{2\pi }$}\psi $  and $\Delta\hat{\mathbf{q}}\ \Delta\hat{\mathbf{p}}\geq \mbox{\small $\frac{d}{4\pi }$}$ are satisfied with approximation for certain  states $\psi_{C,k}$.
}
\begin{tabular}{crr}
\hline
$\psi$ & $||\left([\hat{\mathbf{q}},\hat{\mathbf{p}}]-{\rm i}\frac{d}{2\pi }\right)\psi||$ \qquad & \qquad $|\ |\langle \psi |[\hat{\mathbf{q}},\hat{\mathbf{p}}]|\psi\rangle |-\frac{d}{2\pi }\ |$ \\
\hline
$\psi_{C,0}$ & $2.68152\!\times\!10^{-9}$ & $8.88178\!\times\!10^{-16}$ \\
$\psi_{C,1}$ & $3.67539\!\times\!10^{-8}$ & 0\qquad \mbox{}\\
$\psi_{C,2}$ & $4.93488\!\times\!10^{-7}$ & $5.32907\!\times\!10^{-15}$ \\
$\psi_{C,3}$ & $2.24984\!\times\!10^{-6}$ & $8.88178\!\times\!10^{-15}$\\
$\psi_{C,4}$ & $9.70535\!\times\!10^{-6}$ & $2.67342\!\times\!10^{-13}$\\
$\psi_{C,5}$    & 0.000035908    & $3.03935\!\times\!10^{-12}$\\
$\psi_{C,6}$   & 0.000125868  &  $3.62173\!\times\!10^{-11}$\\
$\psi_{C,7}$    & 0.000281649    & $1.07889\!\times\!10^{-10}$\\
$\psi_{C,8}$   &0.000740305   &  $5.47327\!\times\!10^{-10}$\\
$\psi_{C,9}$    &0.00153664     & $1.43238\!\times\!10^{-9}$\\
$\psi_{C,10}$   &0.00357292   &  $1.30984\!\times\!10^{-10}$\\
\hline
 \end{tabular}
\end{table}
%
%
%
\subsection{Discrete version of the harmonic oscillator}
The operator $\hat{\mathbf{H}}\!:\!\mathcal{H}\!\rightarrow\!\mathcal{H}$,
\begin{equation}\label{DHO}
\hat{\mathbf{H}}\!=\!\frac{\hat{\mathbf{p}}^2\!+\!\hat{\mathbf{q}}^2}{2}
\end{equation}
can be regarded as a discrete version of the Hamiltonian of the quantum harmonic oscillator
\begin{equation}\label{CHO}
\hat{H}\!=\!\frac{\hat{p}^2\!+\!\hat{q}^2}{2}.
\end{equation}
For any small $\varepsilon \!>\!0$, some of the eigenstates of $\hat{\mathbf{H}}$
belong to $\mathcal{S}_\varepsilon$. For example (see Table \ref{HarmonicOsc}),
in the case $d\!=\!31$, the eigenstates  $\psi_{O,0}$, $\psi_{O,1}$, ... , $\psi_{O,7}$ belong to $\mathcal{S}_{0.001}$.
They form an orthonormal system. 
The relation $[\hat{\mathbf{q}},\hat{\mathbf{p}}]\psi\approx \mathrm{i}\frac{d}{2\pi}\psi$ is satisfied if and only if 
$[\mathbf{a},\mathbf{a}^\dag]\psi \approx \psi$. In this case 
$\frac{\hat{\mathbf{p}}^2\!+\!\hat{\mathbf{q}}^2}{2}\psi\!\approx\!(\hat{\mathbf{a}}^\dag \hat{\mathbf{a}}\!+\!\frac{1}{2})\psi$,
and the eigenvalues of $\hat{\mathbf{a}}^\dag \hat{\mathbf{a}}\!+\!\frac{1}{2}$  tend to be 
distributed equidistant,
\begin{equation}
(\hat{\mathbf{a}}^\dag \hat{\mathbf{a}}\!+\!\frac{1}{2})\psi\!=\!\lambda \psi \quad\Longrightarrow\quad 
(\hat{\mathbf{a}}^\dag \hat{\mathbf{a}}\!+\!\frac{1}{2})\hat{\mathbf{a}}^\dag \psi\!=\!
\hat{\mathbf{a}}^\dag (\hat{\mathbf{a}}^\dag \hat{\mathbf{a}}\!+\!\frac{1}{2})\!+\!\hat{\mathbf{a}}^\dag\psi\!=\!(\lambda \!+\!1)\hat{\mathbf{a}}^\dag \psi .
\end{equation}
Therefore, in the bottom of the spectrum of $\frac{\hat{\mathbf{p}}^2\!+\!\hat{\mathbf{q}}^2}{2}$,
the eigenvalues have an almost equidistant distribution (see  Fig. 4).

%
\begin{figure}[h]
\includegraphics[scale=1.2]{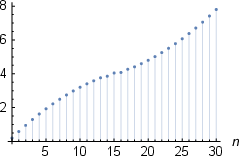}
\caption{Eigenvalues of the Hamiltonian (\ref{DHO})  in the case $d\!=\!31$. }
\end{figure}

%
 \begin{table}[h]
\caption{\label{HarmonicOsc} In the case  $d\!=\!31$, the relations $[\hat{\mathbf{q}},\hat{\mathbf{p}}]\psi \approx{\rm i}\mbox{\small $\frac{d}{2\pi }$}\psi $  and $\Delta\hat{\mathbf{q}}\ \Delta\hat{\mathbf{p}}\geq \mbox{\small $\frac{d}{4\pi }$}$ are satisfied for certain states $\psi_{O,k}$.
}
\begin{tabular}{crr}
\hline
$\psi$ & $||\left([\hat{\mathbf{q}},\hat{\mathbf{p}}]-{\rm i}\frac{d}{2\pi }\right)\psi||$\qquad  & \qquad $|\ |\langle \psi |[\hat{\mathbf{q}},\hat{\mathbf{p}}]|\psi\rangle |-\frac{d}{2\pi }\ |$ \\
\hline
 $\psi_{O,0}$ & $2.73402\!\times\!10^{-9}$ & $5.32907\!\times\!10^{-15}$ \\
 $\psi_{O,1}$ & $2.31647\!\times\!10^{-8}$ & $1.77636\!\times\!10^{-15}$ \\
 $\psi_{O,2}$ & $3.54883\!\times\!10^{-7}$ & $4.44089\!\times\!10^{-15}$ \\
  $\psi_{O,3}$ & $4.8082\!\times\!10^{-7}$ & $1.15463\!\times\!10^{-14}$ \\
  $\psi_{O,4}$ & $5.27306\!\times\!10^{-6}$ & $4.70735\!\times\!10^{-13}$ \\
  $\psi_{O,5}$ & $0.000018252$ & $4.30322\!\times\!10^{-12}$ \\
$\psi_{O,6}$ & $0.000167857$ & $1.50534\!\times\!10^{-10}$ \\
$\psi_{O,7}$ & $0.000127787$ & $7.24515\!\times\!10^{-10}$ \\
$\psi_{O,8}$ & $0.0010158$ & $1.59893\!\times\!10^{-8}$ \\
$\psi_{O,9}$ & $0.00233029$ & $6.25305\!\times\!10^{-8}$ \\
$\psi_{O,10}$ & $0.0176262$ & $1.48606\!\times\!10^{-6}$ \\
\hline
 \end{tabular}
\end{table}
%
%
%
\subsection{Discrete coherent states}
The  {\em discrete coherent states}  \cite{Vourdas04,Cotfas11} are
\begin{equation}\label{discretecs}
\begin{array}{l}
|n,k\rangle \!=\!\frac{1}{\sqrt{d}}\mathbf{D}(n,k)|\mathbf{g}\rangle,
\end{array}
\end{equation}
where $ n,k\!\in\!\{-\!s,\!-s\!+\!1,...,s\!-\!1,s\}$ and 
$\mathbf{D}(n ,k )\!:\!\mathcal{H}\!\rightarrow\!\mathcal{H},$
\begin{equation}
\begin{array}{l}
 \mathbf{D}(n ,k )\psi (m)=\mathrm{e}^{-\frac{\pi \mathrm{i}}{d}nk}\, \mathrm{e}^{\frac{2\pi \mathrm{i}}{d}km}\, \psi (m\!-\!n ),
\end{array}
\end{equation}
are the {\em displacement operators}. By using the relations 
\begin{equation}
\parallel \!\mathbf{C}|n,k\rangle \!\parallel^2=\!\!\!
\sum\limits_{m=-s}^s\left| \frac{1}{d}\right. \!\!\sum\limits_{j=-s}^s\sum\limits_{\ell=-s}^s\!\!
j(m\!-\!\ell)\, \mathrm{e}^{\frac{2\pi \mathrm{i}}{d}j(m-\ell)} 
\ \mathrm{e}^{\frac{2\pi \mathrm{i}}{d}k\ell }\,  \mathbf{g}(\ell\!-\!n)
\left.-\, \mathrm{i}\frac{d}{2\pi}\ \mathrm{e}^{\frac{2\pi \mathrm{i}}{d}km}\ \mathbf{g}(m\!-\!n)\right|^2,
\end{equation}
$|\bar{z}|\!=\!|z|$,\ 
$\mathbf{g}(-n)\!=\!\mathbf{g}(n)$ and the changes of parameters
$n\!\mapsto \!-n$, $k\!\mapsto \!-k$, $m\!\mapsto \!-m$, one can prove that
\begin{equation}
\parallel \!\mathbf{C}|n,k\rangle \!\!\parallel=\parallel \!\mathbf{C}|-n,k\rangle \!\! \parallel=\parallel \!\mathbf{C}|n,-k\rangle \!\!\parallel=\parallel \!\mathbf{C}|-n,-k\rangle \!\!\parallel.
\end{equation}

For any small $\varepsilon \!>\!0$, some of the discrete coherent states $|j,k\rangle $
belong to $\mathcal{S}_\varepsilon$ if the Hilbert space of the quantum system has a  dimension high enough. For example (see Table \ref{discrcoh}),
in the case $d\!=\!31$, the coherent states  
$|0,0\rangle $, $|0,\pm 1\rangle $, $|0,\pm 2\rangle $, $|0,\pm 3\rangle $, $|0,\pm 4\rangle $, 
$|\pm 1,0\rangle $, $|\pm 1,\pm 1\rangle $, $|\pm 1,\pm 2\rangle $, $|\pm 1,\pm 3\rangle $, $|\pm 1,\pm 4\rangle $, 
$|\pm 2,0\rangle $, $|\pm 2,\pm 1\rangle $, $|\pm 2,\pm 2\rangle $, $|\pm 2,\pm 3\rangle $, $|\pm 2,\pm 4\rangle $, 
$|\pm 3,0\rangle $, $|\pm 3,\pm 1\rangle $, $|\pm 3,\pm 2\rangle $, $|\pm 3,\pm 3\rangle $, $|\pm 3,\pm 4\rangle $, 
$|\pm 4,0\rangle $, $|\pm 4,\pm 1\rangle $, $|\pm 4,\pm 2\rangle $, $|\pm 4,\pm 3\rangle $, $|\pm 4,\pm 4\rangle $ 
belong to $\mathcal{S}_{0.001}$.
 \begin{table*}[h]
\caption{\label{discrcoh} In the case  $d\!=\!31$, the relations $[\hat{\mathbf{q}},\hat{\mathbf{p}}]\psi \approx{\rm i}\mbox{\small $\frac{d}{2\pi }$}\psi $  and $\Delta\hat{\mathbf{q}}\ \Delta\hat{\mathbf{p}}\geq \mbox{\small $\frac{d}{4\pi }$}$ are satisfied with approximation for certain states $|j,k\rangle $.
}
\begin{tabular}{cccccc}
\hline
$|n,k\rangle$ \qquad & \quad  $||\left([\hat{\mathbf{q}},\hat{\mathbf{p}}]-{\rm i}\frac{d}{2\pi }\right)|n,k\rangle||$ \quad  & \qquad  $|\ |\langle n,k |[\hat{\mathbf{q}},\hat{\mathbf{p}}]|n,k\rangle |-\frac{d}{2\pi }\ |$  \qquad \qquad  &   $|n,k\rangle$  \qquad   & \quad \ \  $||\left([\hat{\mathbf{q}},\hat{\mathbf{p}}]-{\rm i}\frac{d}{2\pi }\right)|n,k\rangle||$  \qquad   & \ \   $|\ |\langle n,k |[\hat{\mathbf{q}},\hat{\mathbf{p}}]|n,k\rangle |-\frac{d}{2\pi }\ |$ \ \  \\
\hline
$|0,0\rangle$ &  $2.68152\!\times\!10^{-9}$ & $8.88178\!\times\!10^{-16}$ & $|3,0\rangle$ & $8.88946\!\times\!10^{-6}$ & $2.70894\!\times\!10^{-13}$\\
$|0,1\rangle$ &  $3.53522\!\times\!10^{-8}$ & $1.77636\!\times\!10^{-15}$ & $|3,1\rangle$ & $8.88748\!\times\!10^{-6}$ & $2.70894\!\times\!10^{-13}$\\
$|0,2\rangle$ &  $6.06924\!\times\!10^{-7}$ & $8.88178\!\times\!10^{-16}$ & $|3,2\rangle$ & $9.21918\!\times\!10^{-6}$ & $2.80664\!\times\!10^{-13}$\\
$|0,3\rangle$ & $8.88946\!\times\!10^{-6}$   & $2.73559\!\times\!10^{-13}$ & $|3,3\rangle$ & $0.000011576$ & $5.21361\!\times\!10^{-13}$\\
$|0,4\rangle$ & $0.000105793$   & $4.01679\!\times\!10^{-11}$ & $|3,4\rangle$ & $0.000107044$ & $4.00791\!\times\!10^{-11}$\\
$|0,5\rangle$ & $0.00102558$   & $3.9099\!\times\!10^{-9}$ & $|3,5\rangle$ & $0.00104948$ & $3.91981\!\times\!10^{-9}$\\
$|1,0\rangle$ &  $3.53522\!\times\!10^{-8}$ & $8.88178\!\times\!10^{-16}$ & $|4,0\rangle$ & $0.000105793$ & $4.01696\!\times\!10^{-11}$\\
$|1,1\rangle$ &  $3.53347\!\times\!10^{-8}$ & $2.66454\!\times\!10^{-15}$ & $|4,1\rangle$ & $0.000106023$ & $4.01714\!\times\!10^{-11}$\\
$|1,2\rangle$ &  $6.2644\!\times\!10^{-7}$ & $0$ & 
$|4,2\rangle$ & $0.000106471$ & $4.01306\!\times\!10^{-11}$\\
$|1,3\rangle$ & $8.88748\!\times\!10^{-6}$   & $2.74447\!\times\!10^{-13}$ & 
$|4,3\rangle$ & $0.000107044$ & $4.00791\!\times\!10^{-11}$\\
$|1,4\rangle$ & $0.000106023$   & $4.01732\!\times\!10^{-11}$ & 
$|4,4\rangle$ & $0.000186876$ & $9.42757\!\times\!10^{-11}$\\
$|1,5\rangle$ & $0.00102781$   & $3.90986\!\times\!10^{-9}$ & 
$|4,5\rangle$ & $0.00100292$ & $3.78383\!\times\!10^{-9}$\\
$|2,0\rangle$ &  $6.06924\!\times\!10^{-7}$ & $1.77636\!\times\!10^{-15}$ & $|5,0\rangle$ & $0.00102558$ & $3.9099\!\times\!10^{-9}$\\
$|2,1\rangle$ &  $6.2644\!\times\!10^{-7}$ & $8.88178\!\times\!10^{-16}$ & $|5,1\rangle$ & $0.00102781$ & $3.90986\!\times\!10^{-9}$\\
$|2,2\rangle$ &  $6.28767\!\times\!10^{-7}$ & $2.66454\!\times\!10^{-15}$ & 
$|5,2\rangle$ & $0.00103463$ & $3.91002\!\times\!10^{-9}$\\
$|2,3\rangle$ & $9.21918\!\times\!10^{-6}$   & $2.77112\!\times\!10^{-13}$ & 
$|5,3\rangle$ & $0.00104948$ & $3.91981\!\times\!10^{-9}$\\
$|2,4\rangle$ & $0.000106471$   & $4.01323\!\times\!10^{-11}$ & 
$|5,4\rangle$ & $0.00100292$ & $3.78383\!\times\!10^{-9}$\\
$|2,5\rangle$ & $0.00103463$   & $3.91002\!\times\!10^{-9}$ & 
$|5,5\rangle$ & $0.00170788$ & $8.59234\!\times\!10^{-9}$\\
\hline
\end{tabular}
\end{table*}
%
%
%
\subsection{Eigenfunctions of discrete annihilation operator}
It is known that, in the continuous-variable case, the standard coherent states 
$|q,p\rangle$ can be defined as eigenstates of the annihilation operator, $\hat{a}|\alpha\rangle\!=\!\alpha |\alpha\rangle$, that is
\begin{equation}
\begin{array}{l}
\hat{a}\, |q,p\rangle\!=\!\sqrt{\frac{\pi }{h}}(q\!+\!\mathrm{i}p)|q,p\rangle .
\end{array}
\end{equation}
By following the analogy, we consider the eigenstates $\psi_{A,n}$ of the discrete version 
\begin{equation}
\hat{\mathbf{a}}\!=\!\sqrt{\frac{\pi}{d}}(\hat{\mathbf{q}}\!+\!{\rm i}\hat{\mathbf{p}})
\end{equation}
of $\hat{a}$.
For any small $\varepsilon \!>\!0$, some of the eigenfunctions  $\psi_{A,n} $
belong to $\mathcal{S}_\varepsilon$ if the Hilbert space of the quantum system has a  dimension high enough. For example (see Table \ref{AAA}),
in the case $d\!=\!31$, the eigenstates  
$\psi_{A,0}$, $\psi_{A,1}$, ... , $\psi_{A,14}$, where the states have been considered in the increasing order of the number of sign oscillations, 
belong to $\mathcal{S}_{0.001}$.

 \begin{table}{h}
\caption{ \label{AAA} In the case $d\!=\!31$, the relations $[\hat{\mathbf{q}},\hat{\mathbf{p}}]\psi \approx{\rm i}\mbox{\small $\frac{d}{2\pi }$}\psi $  and $\Delta\hat{\mathbf{q}}\ \Delta\hat{\mathbf{p}}\geq \mbox{\small $\frac{d}{4\pi }$}$ are satisfied with approximation by certain  states $\psi_{A,k}$.
}
\begin{tabular}{crrlrr}
\hline
$\psi$ & $||\left([\hat{\mathbf{q}},\hat{\mathbf{p}}]-{\rm i}\frac{d}{2\pi }\right)\psi||$ & \qquad $|\ |\langle \psi |[\hat{\mathbf{q}},\hat{\mathbf{p}}]|\psi\rangle |-\frac{d}{2\pi }\ |$ &
\qquad \ \ $\psi$ & $||\left([\hat{\mathbf{q}},\hat{\mathbf{p}}]-{\rm i}\frac{d}{2\pi }\right)\psi||$ & \qquad $|\ |\langle \psi |[\hat{\mathbf{q}},\hat{\mathbf{p}}]|\psi\rangle |-\frac{d}{2\pi }\ |$
\\
\hline
$\psi_{A,0}$ & $2.08125\!\times\!10^{-9}$ & 0\qquad \mbox{} & 
\qquad $\psi_{A,10}$ & $3.2588\!\times\!10^{-6}$ & $2.75335\!\times\!10^{-14}$\\
$\psi_{A,1}$ & $2.08125\!\times\!10^{-9}$ & $8.88178\!\times\!10^{-16}$ &
\qquad $\psi_{A,11}$    & 0.000746798    & $1.2461\!\times\!10^{-11}$\\
$\psi_{A,2}$ & $2.08126\!\times\!10^{-9}$ & $1.77636\!\times\!10^{-15}$ & 
\qquad $\psi_{A,12}$    & 0.000746798    & $1.2461\!\times\!10^{-11}$\\
$\psi_{A,3}$ & $1.41314\!\times\!10^{-7}$ & $8.88178\!\times\!10^{-16}$ & 
\qquad $\psi_{A,13}$    & 0.000746798    & $1.2461\!\times\!10^{-11}$\\
$\psi_{A,4}$ & $1.41314\!\times\!10^{-7}$ & $8.88178\!\times\!10^{-16}$ & 
\qquad $\psi_{A,14}$    & 0.000746798    & $1.2461\!\times\!10^{-11}$\\
$\psi_{A,5}$ & $1.41314\!\times\!10^{-7}$ & $1.77636\!\times\!10^{-15}$ & 
\qquad $\psi_{A,15}$    & 0.00169824    & $5.93746\!\times\!10^{-9}$\\
$\psi_{A,6}$ & $1.41314\!\times\!10^{-7}$ & 0\qquad \mbox{}& 
\qquad $\psi_{A,16}$    & 0.00169824    & $5.93746\!\times\!10^{-9}$\\
$\psi_{A,7}$ & $3.2588\!\times\!10^{-6}$ & $2.57572\!\times\!10^{-14}$ & \qquad $\psi_{A,17}$    & 0.00169824    & $5.93746\!\times\!10^{-9}$
\\
$\psi_{A,8}$ & $3.2588\!\times\!10^{-6}$ & $2.66454\!\times\!10^{-14}$ & \qquad $\psi_{A,18}$    & 0.00169824    & $5.93746\!\times\!10^{-9}$\\
$\psi_{A,9}$ & $3.2588\!\times\!10^{-6}$ & $2.57572\!\times\!10^{-14}$  & \qquad  $\psi_{A,19}$   & 0.0381896  &  $2.72907\!\times\!10^{-6}$\\
\hline
 \end{tabular}
\end{table}
%
%
%
\subsection{Discrete coherent state quantization}
In the continuous case the standard coherent states 
\begin{equation}
|q,p\rangle \!=\!D(q,p)|g\rangle,
\end{equation}
where 
\begin{equation}
\begin{array}{l}
g(q)\!=\!\frac{1}{\sqrt[4]{\pi}} \,\mathrm{e}^{-\frac{1}{2}q^2}
\end{array}
\end{equation}
is the vacuum state, satisfy the decomposition of the identity
\begin{equation}
\begin{array}{l}
\frac{1}{2\pi}\int\!\!\!\int\limits_{\mathbb{R}^2}|q,p\rangle \langle q,p|\, dqdp\!=\!\mathbb{I}_{L^2(\mathbb{R})}.
\end{array}
\end{equation}
To any function $f\!:\!\mathbb{R}^2\!\rightarrow\mathbb{C}$ defined on the phase space $\mathbb{R}^2$, and chosen such that the integral is convergent, one can 
associate the operator
\begin{equation}
\begin{array}{l}
A_f\!=\!\frac{1}{2\pi}\int\!\!\!\int\limits_{\mathbb{R}^2}f(q,p)\, |q,p\rangle \langle q,p|\, dqdp.
\end{array}
\end{equation}
The map $f\!\mapsto A_f$ is called coherent state quantization \cite{Cotfas11,Gazeau,Perelomov}
. One can prove that \cite{Gazeau}
\begin{equation}
\begin{array}{lll}
A_f\!=\!\hat{q} & \mbox{for} & f(q,p)=q,\\
A_f\!=\!\hat{p} & \mbox{for} & f(q,p)=p,\\
A_f\!=\!\frac{1}{2}(\hat{p}^2\!+\!\hat{q}^2)\!+\!\frac{1}{2} & \mbox{for} & f(q,p)=\frac{1}{2}({p}^2\!+\!{q}^2).
\end{array}
\end{equation}

Because the discrete coherent states satisfy 
\begin{equation}
\sum\limits_{n,k=-s}^s|n,k\rangle \langle n,k|\!=\!\mathbb{I},
\end{equation}
by following the analogy with the continuous case, one can associate the operator
$\mathbf{A}_f\!:\!\mathcal{H}\!\rightarrow \!\mathcal{H}$,
\begin{equation}
\begin{array}{l}
\mathbf{A}_f\!=\!\sum\limits_{n,k=-s}^s f(n,k)\, |n,k\rangle \langle n,k|
\end{array}
\end{equation}
to each function
\begin{equation}
f\!:\!\{-s,-s\!+\!1,...,s\}\!\times\!\{-s,-s\!+\!1,...,s\}\!\rightarrow \mathbb{C} 
\end{equation}
defined on the discrete phase space.
By using the discrete coherent state quantization  $f\!\mapsto\!\mathbf{A}_f$, 
we obtain for $f(n,k)\!=\!\frac{1}{2}(n^2\!+\!k^2)$ the operator
\begin{equation}\label{Hamq}
\begin{array}{l}
\mathbf{H}_q\!=\!-\frac{1}{2}\!+\!\sum\limits_{n,k=-s}^s \frac{n^2\!+\!k^2}{2}\, |n,k\rangle \langle n,k|
\end{array}
\end{equation}
which can be regarded as a discrete version of the harmonic oscillator Hamiltonian.
For any small $\varepsilon \!>\!0$, some of the eigenfunctions  $\psi_{Q,n} $ of $\mathbf{H}_q$ 
belong to $\mathcal{S}_\varepsilon$ if the Hilbert space of the quantum system has a  dimension high enough. For example (see Table \ref{coherquant}),
in the case $d\!=\!31$, the states  
$\psi_{Q,0}$, $\psi_{Q,1}$, ... , $\psi_{Q,7}$, where the states have been considered in the increasing order of eigenvalues, 
belong to $\mathcal{S}_{0.001}$. One can again remark an almost equidistant distribution in the bottom of the spectrum (see  Fig. 5). The Hamiltonian (\ref{Hamq}) is an other discrete version of 
(\ref{CHO}), and a behavior similar to (\ref{DHO}) is again expected to happen.

 \begin{table}[h]
\caption{\label{coherquant} In the case $d\!=\!31$, the relations $[\hat{\mathbf{q}},\hat{\mathbf{p}}]\psi \approx{\rm i}\mbox{\small $\frac{d}{2\pi }$}\psi $  and $\Delta\hat{\mathbf{q}}\ \Delta\hat{\mathbf{p}}\geq \mbox{\small $\frac{d}{4\pi }$}$ are satisfied with approximation of certain  states $\psi_{Q,k}$.
}
\begin{tabular}{crr}
\hline
$\psi$ & $||\left([\hat{\mathbf{q}},\hat{\mathbf{p}}]-{\rm i}\frac{d}{2\pi }\right)\psi||$ & \qquad $|\ |\langle \psi |[\hat{\mathbf{q}},\hat{\mathbf{p}}]|\psi\rangle |-\frac{d}{2\pi }\ |$ \\
\hline
 $\psi_{Q,0}$ & $3.95539\!\times\!10^{-9}$ & $1.77636\!\times\!10^{-15}$ \\
 $\psi_{Q,1}$ & $3.45774\!\times\!10^{-8}$ & $2.66454\!\times\!10^{-15}$ \\
 $\psi_{Q,2}$ & $5.48854\!\times\!10^{-7}$ & $4.44089\!\times\!10^{-15}$ \\
  $\psi_{Q,3}$ & $6.77894\!\times\!10^{-7}$ & $2.4869\!\times\!10^{-14}$ \\
  $\psi_{Q,4}$ & $7.94567\!\times\!10^{-9}$ & $9.82325\!\times\!10^{-13}$ \\
  $\psi_{Q,5}$ & $0.0000278873$ & $8.32134\!\times\!10^{-12}$ \\
$\psi_{Q,6}$ & $0.00027139$ & $3.52218\!\times\!10^{-10}$ \\
$\psi_{Q,7}$ & $0.000187143$ & $1.42712\!\times\!10^{-9}$ \\
$\psi_{Q,8}$ & $0.00162965$ & $3.83563\!\times\!10^{-8}$ \\
$\psi_{Q,9}$ & $0.00376062$ & $1.33373\!\times\!10^{-7}$ \\
$\psi_{Q,10}$ & $0.0310428$ & $4.15222\!\times\!10^{-6}$ \\
\hline
 \end{tabular}
\end{table}
%
%
\begin{figure}[h]
\includegraphics[scale=1.2]{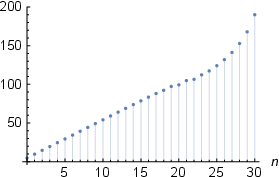}
\caption{Eigenvalues of the Hamiltonian (\ref{Hamq})  in the case $d\!=\!31$. }
\end{figure}
%
%
%
\subsection{Mehta states}
It is known that, in the continuous-variable case, the eigenstates of the quantum 
harmonic oscillator $\hat{H}\!=\!\frac{\hat{p}^2\!+\!\hat{q}^2}{2}$ can be defined
in terms of Hermite polynomials as $\Psi_n\!:\!(-\infty,\infty)\!\rightarrow \mathbb{R}$,
\begin{equation}
\begin{array}{l}
\Psi_n(q)\!=\!\frac{1}{\sqrt{n!\, 2^n\sqrt{\pi}}}\, H_n(q)\ \mathrm{e}^{-\frac{1}{2}q^2}
\end{array}
\end{equation}
and satisfy the relation 
\begin{equation}
\begin{array}{l}
\hat{H}\Psi_n\!=\!(n\!+\!\frac{1}{2})\Psi_n.
\end{array}
\end{equation}
It is known that the Hermite-Gauss functions $\Psi_n$ are also eigenfunctions
of the Fourier transform, namely
\begin{equation}
F\Psi_n\!=\!(-\mathrm{i})^n\Psi_n.
\end{equation}

In his article \cite{Mehta87}, Mehta has defined the discrete version 
$\psi_{M,n}\!:\!\{-s,-s\!+\!1,...,s\}\!\rightarrow\!\mathbb{R}$,
\begin{equation}
\begin{array}{l}
\psi_{M,n}(m)\!=\!\sum\limits_{\alpha=-\infty}^\infty H_n(m\!+\!\alpha d)\ 
\mathrm{e}^{-\frac{1}{2}(m\!+\!\alpha d)^2}
\end{array}
\end{equation}
of $\Psi_n$, and proved that it is an eigenfunction of the discrete Fourier transform,
\begin{equation}
\mathbf{F}\psi_{M,n}\!=\!(-\mathrm{i})^n\psi_{M,n}.
\end{equation}
For any small $\varepsilon \!>\!0$, some states $\psi_{M,n} $
belong to $\mathcal{S}_\varepsilon$ if the Hilbert space of the quantum system has a  dimension high enough. For example (see Table \ref{MMM}),
in the case $d\!=\!31$, the functions  
$\psi_{M,0}$, $\psi_{M,1}$, ... , $\psi_{M,9}$,
after normalization,  belong to $\mathcal{S}_{0.001}$.

 \begin{table}[h]
\caption{ \label{MMM}
In the case $d\!=\!31$, the relations $[\hat{\mathbf{q}},\hat{\mathbf{p}}]\psi \approx{\rm i}\mbox{\small $\frac{d}{2\pi }$}\psi $  and $\Delta\hat{\mathbf{q}}\ \Delta\hat{\mathbf{p}}\geq \mbox{\small $\frac{d}{4\pi }$}$ are satisfied with approximation of certain Mehta states $\psi_{M,k}$.
}
\begin{tabular}{crr}
\hline
$\psi$ & \qquad $||\left([\hat{\mathbf{q}},\hat{\mathbf{p}}]-{\rm i}\frac{d}{2\pi }\right)\psi||$ & \qquad  $|\ |\langle \psi |[\hat{\mathbf{q}},\hat{\mathbf{p}}]|\psi\rangle |-\frac{d}{2\pi }\ |$ \\
\hline
$\psi_{M,0}$ & $2.68152\!\times\!10^{-9}$ &  $8.88178\!\times\!10^{-16}$\\
$\psi_{M,1}$ &$2.41961\!\times\!10^{-8}$ & 0\qquad \mbox{}\\
$\psi_{M,2}$ &$2.98432\!\times\!10^{-7}$ & $2.66454\!\times\!10^{-15}$\\
$\psi_{M,3}$ & $349482\!\times\!10^{-7}$ &$1.1543\!\times\!10^{-14}$\\
$\psi_{M,4}$ & $2.95378\!\times\!10^{-6}$    & $1.55431\!\times\!10^{-13}$\\
$\psi_{M,5}$ & $8.69175\!\times\!10^{-6}$ &  $9.28146\!\times\!10^{-13}$\\
$\psi_{M,6}$ & 0.00005669 & $1.78559\!\times\!10^{-11}$\\
$\psi_{M,7}$ & 0.00003822 & $5.83524\!\times\!10^{-11}$\\
$\psi_{M,8}$ & 0.0002320 &$7.42824\!\times\!10^{-10}$\\
$\psi_{M,9}$ & 0.0004216 &$2.07663\!\times\!10^{-9}$\\
$\psi_{M,10}$ & 0.0022077    & $2.6385\!\times\!10^{-8}$\\
\hline
 \end{tabular}
\end{table}
%
%
\subsection{Harper states}
In continuous-variable case, $\hat{p}\!=\!F^\dag\hat{q}F$, and 
the harmonic oscillator Hamiltonian can be written as
\begin{equation}
\hat{H}\!=\!\frac{\hat{p}^2\!+\!\hat{q}^2}{2}\!=
\!\frac{\hat{p}^2\!+\!F\hat{p}^2F^\dag}{2}.
\end{equation}
A discrete version of the quantum harmonic oscillator can be obtained by using 
for the differential operator $\frac{\mathrm{d}^2}{\mathrm{d}q^2}$ the finite-difference version $D^2\!:\!\mathcal{H}\!\rightarrow\!\mathcal{H}$,
\begin{equation}
D^2\psi(n)\!=\!\psi(n\!-\!1)\!-\!2\psi(n)\!+\!\psi(n\!+\!1)
\end{equation}
 and the relation
\begin{equation}
\begin{array}{l}
\mathbf{F}D^2\mathbf{F}^\dag\psi(n)\!=\!\left(2 \cos\frac{2\pi n}{d}\!-\!1 \right)\psi(n).
\end{array}
\end{equation}
The eigenfunctions $\psi_{H,n}\!:\!\{-s,-s\!+\!1,...,s, s\!+\!1\}\!\rightarrow\!\mathbb{R}$ of 
\begin{equation}\label{Harp}
\hat{\mathbf{H}}_d\!=\!-\frac{1}{2}(D^2\!+\!\mathbf{F}D^2\mathbf{F}^\dag)
\end{equation}
correspond to the normalized eigenvectors of the matrix (addition is modulo $d$)
\begin{equation}
\begin{array}{l}
\left( 2\cos\frac{2\pi n}{d}\delta_{nk}\!-\!4\!+\!\delta_{k,n+1}
\!+\!\delta_{k,n-1}\right)_{\!\!-s\leq n,k\leq s}
\end{array}
\end{equation}
and are called {\em Harper functions} \cite{Barker}.

For any small $\varepsilon \!>\!0$, some states $\psi_{H,n} $
belong to $\mathcal{S}_\varepsilon$ if the Hilbert space of the quantum system has a  dimension high enough. For example (see Table \ref{HHH}),
in the case $d\!=\!31$, the eigenstates  
$\psi_{H,0}$, $\psi_{H,1}$, $\psi_{H,2}$ , $\psi_{H,3}$
belong to $\mathcal{S}_{0.001}$.
Again, there is an almost equidistant distribution in the lower part of the spectrum (see  Fig. 6). The Hamiltonian (\ref{Harp}) is an other discrete version of 
(\ref{CHO}), and a behavior similar to (\ref{DHO}) is again expected to happen.

 \begin{table}[h]
\caption{\label{HHH} In the case $d\!=\!31$, the relations $[\hat{\mathbf{q}},\hat{\mathbf{p}}]\psi \approx{\rm i}\mbox{\small $\frac{d}{2\pi }$}\psi $  and $\Delta\hat{\mathbf{q}}\ \Delta\hat{\mathbf{p}}\geq \mbox{\small $\frac{d}{4\pi }$}$ are satisfied with approximation of certain Harper states $\psi_{H,k}$.
}
\begin{tabular}{crr}
\hline
$\psi$ & $||\left([\hat{\mathbf{q}},\hat{\mathbf{p}}]-{\rm i}\frac{d}{2\pi }\right)\psi||$ & \qquad $|\ |\langle \psi |[\hat{\mathbf{q}},\hat{\mathbf{p}}]|\psi\rangle |-\frac{d}{2\pi }\ |$ \\
\hline
$\psi_{H,0}$ & $1.75165\!\times\!10^{-6}$ & $4.08562\!\times\!10^{-14}$  \\
$\psi_{H,1}$ & 0.0000101661 & $8.56204\!\times\!10^{-13}$ \\
$\psi_{H,2}$ & 0.000218006 & $1.87562\!\times\!10^{-10}$ \\
 $\psi_{H,3}$ & 0.000136764 &  $6.73229\!\times\!10^{-10}$ \\
 $\psi_{H,4}$ & 0.00211415 &  $5.87102\!\times\!10^{-8}$ \\
 $\psi_{H,5}$ & 0.00445289 &  $1.51256\!\times\!10^{-7}$ \\
$\psi_{H,6}$ & 0.0632087 &  $0.0000150119$ \\
$\psi_{H,7}$ & 0.0187208 &  $0.0000122191$ \\
\hline
 \end{tabular}
\end{table}
%
%
\begin{figure}[h]
\includegraphics[scale=1.2]{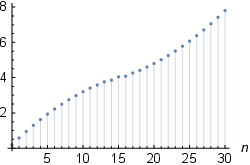}
\caption{Eigenvalues of the Hamiltonian (\ref{Harp})  in the case $d\!=\!31$. }
\end{figure}
%
%
\subsection{Kravchuk states}
The Kravchuk polynomials $K_m$, defined by 
\begin{equation}
(1\!-\!X)^{s+k}(1\!+\!X)^{s-k}\!=\!\sum\limits_{m=-s}^sK_m(k)\, X^{s+m}
\end{equation}
are
\begin{equation}
K_m(k)\!=\!\sum\limits_{n=0}^{s+m}(-1)^n\ C_{s+k}^m\ C_{s-k}^{s+m-n}
\end{equation}
and satisfy the relation
\begin{equation}
\begin{array}{l}
\frac{1}{2^{2s}}\sum\limits_{k=-s}^s C_{2s}^{s+k}\ K_m(k)\ K_n(k)\!=\!
C_{2s}^{s+m}\ \delta_{mn},
\end{array}
\end{equation}
where
\begin{equation}
C_m^n
\!=\!\left\{
\begin{array}{cll}
\frac{m!}{n!\ (m\!-\!n)!} & \mbox{for} & n\!\in\!\{0,1,...,m\}\\
0 & \mbox{for} & n\!\not\in\!\{0,1,...,m\}.
\end{array}\right.
\end{equation}
The Kravchuk functions \cite{Nikiforov,Vilenkin}
\[
\psi_{K,-s},\ \psi_{K,-s+1},\, ...\, \ \psi_{K,s}:\{ -s,-s\!+\!1,...,s\!-\!1,s\}\!\rightarrow\!\mathbb{R},
\]
\begin{equation}
\psi_{K,m}(n)\!=\!\frac{1}{2^s}\sqrt{\frac{C_{2s}^{s+n}}{ C_{2s}^{s+m}}}\  K_m(n),
\end{equation}
form an orthonormal basis in $\mathcal{H}$.

For any small $\varepsilon \!>\!0$, some states $\psi_{K,n} $
belong to $\mathcal{S}_\varepsilon$ if the Hilbert space of the quantum system has a  dimension high enough. For example (see Table \ref{KKK}),
in the case $d\!=\!31$, the eigenstates  
$\psi_{K,-15}$, $\psi_{K,-14}$, $\psi_{K,-13}$,
belong to $\mathcal{S}_{0.1}$.
 \begin{table}[h]
\caption{\label{KKK} In the case $d\!=\!31$, the relations $[\hat{\mathbf{q}},\hat{\mathbf{p}}]\psi \approx{\rm i}\mbox{\small $\frac{d}{2\pi }$}\psi $  and $\Delta\hat{\mathbf{q}}\ \Delta\hat{\mathbf{p}}\geq \mbox{\small $\frac{d}{4\pi }$}$ are satisfied with approximation of certain Kravchuk states $\psi_{K,k}$.
}
\begin{tabular}{crr}
\hline
$\psi$ & $||\left([\hat{\mathbf{q}},\hat{\mathbf{p}}]-{\rm i}\frac{d}{2\pi }\right)\psi||$ & \qquad $|\ |\langle \psi |[\hat{\mathbf{q}},\hat{\mathbf{p}}]|\psi\rangle |-\frac{d}{2\pi }\ |$ \\
\hline
$\psi_{K,-15}$ &0.00188 & $3.3511\!\times\!10^{-8}$\\
$\psi_{K,-14}$ &0.010478 & $1.06622\!\times\!10^{-6}$\\
$\psi_{K,-13}$ & 0.0435235 & $0.0000165$\\
\hline
\end{tabular}
\end{table}

%
%
\section{Some possible physical interpretations}
We have proved, mainly numerically, that in the case of a 
quantum system (qudit) described by a $d$-dimensional Hilbert space,
a position-like operator $\hat{\mathbf{q}}$ and a momentum-like operator $\hat{\mathbf{q}}$ satisfy $[\hat{\mathbf{q}},\hat{\mathbf{p}}]\psi\approx \mathrm{i}\frac{d}{2\pi}\psi$ for certain states $\psi$ if the dimension $d$ is high enough. A fundamental question concerns the existence of any physical interpretation for the operators $\hat{\mathbf{q}}$ and $\hat{\mathbf{p}}$ defined mathematically by following the analogy with the continuous-variable case.
It is known that, in the case of a quantum system with $SU(2)$ symmetry, the spectrum of $J_z$ has the form $\{-s,-s\!+\!1,..,s\!-\!1,s\}$. So, for example, we can choose $\hat{\mathbf{q}}\!=\!J_z$
and $\hat{\mathbf{p}}$ to be the Fourier conjugate operator.
More general, any observable with equidistant eigenvalues can play the role of $\hat{\mathbf{q}}$. For eigenvectors forming a basis of the Hilbert space there exists a privileged order, namely the order for which the corresponding eigenvalues form an increasing sequence.

The appearance of quantum states which satisfy with approximation the discrete version of the position-momentum commutation relation among the eigenstates with small energy of certain Hamiltonians suggests a possibility to experimentally obtain such states. They occur naturally when we measure a discrete version of the harmonic quantum
oscillator Hamiltonian.


%
%
\section{Concluding remarks}
In the case of a continuous-variable quantum system, the canonical position-momentum commutation relation is
exactly satisfied in a subspace of the Hilbert space, and plays a fundamental role.
The discrete version of this important relation is also satisfied in a subspace, but not exactly, and this happens only if the dimension
of the Hilbert space is high enough.
We have investigated the behavior with respect to this discrete version of  some important states used in the mathematical description of qudits, including discrete Gaussian states, discrete coherent states and discrete versions of the Hermite-Gauss states.
Thus, we have obtained a new property (to our knowledge)
of these important quantum states. The discrete version of the position-momentum is satisfied with approximation of certain of them if the dimension of the Hilbert space is high enough.
The numerical data presented in this article (Figs. 1-6 and Tables I-XIII) have been obtained using the computer programs presented at the end of the preprint \cite{Cotfas-arXiv}.

At this moment, there is no satisfactory mathematical explanation for the behavior of the analyzed quantum states 
concerning the discrete version of the position-momentum 
commutation relation. Why there exist states for which 
the relation is satisfied with approximation ? Why they are among 
the the low energy states ? Why their existence depends on the
dimension of the Hilbert space ? There seems to exist a
connection between the approximate satisfaction of the  position-momentum commutation relation and the tendency of corresponding eigenvalues to be distributed equidistant.

Quantum Information uses a continuous-variable description, but
it is mainly limited to a very particular class of states (including the Gaussian states) and a very particular class of unitary transforms (including the Gaussian unitaries).
The discrete-variable version of the commutation relation of  position -momentum operators is acceptable satisfied by certain particular states.
We think that, some useful models involving only such particular states, can be built in the discrete-variable case by following the analogy with the continuous-variable case. 

Qudit-based quantum computing \cite{Wang2020} is an alternative to the conventional qubit-based quantum computing. The use of quantum systems described by 
Hilbert spaces of dimension $d\!>\!2$ provides a larger space to store and process information, and the ability to do multiple control operations simultaneously. These features may allow a reduction of
the circuit complexity, a simplification of the experimental setup, and offer the possibility to use more efficient algorithms.


%
%
\mbox{}\\[5cm]
\noindent{\bf The numerical data have been obtained by using the following computer programs in {\small MATHEMATICA}}\\[3mm]
\rule{2mm}{2mm}\ \ {\bf {\small FIGURE} 1}\\[2mm]
Fig.1 has been obtained by using the program:\\[-5mm]
\begin{verbatim}
d = 31; s := (d - 1)/2 ; t = 0
g[alpha_, n_] :=N[Sum[Exp[-Pi  alpha (m d + n + t)^2/d], {m, -Infinity, Infinity}]]
normg[alpha_] := N[Sqrt[Sum[g[alpha, m]^2, {m, -s, s}]]]
gg1 := N[Table[ g[3, n]/normg[3], {n, -s, s}]]
gg2 := N[Table[ g[1/3, n]/normg[1/3], {n, -s, s}]]
ListPlot[Table[{n, gg1[[n + s + 1]]}, {n, -s, s}], Filling -> Axis, 
 AxesLabel -> {n,}, AspectRatio -> 2/3, PlotRange -> Full]
ListPlot[Table[{n, gg2[[n + s + 1]]}, {n, -s, s}], Filling -> Axis, 
 AxesLabel -> {n,}, AspectRatio -> 2/3, PlotRange -> Full]
\end{verbatim}
\rule{2mm}{2mm}\ \ {\bf {\small FIGURE} 2}\\[2mm]
Fig.2 has been obtained by using for $d\!=\!11$ and $d\!=\!31$ the program:\\[-5mm]
\begin{verbatim}
d = 11; s := (d - 1)/2 
g[alpha_, n_] := N[Sum[Exp[-Pi  alpha (m d + n)^2/d], {m, -10, 10}]]
normg[alpha_] := N[Sqrt[Sum[g[alpha, m]^2, {m, -s, s}]]]
psi[alpha_, n_] := g[alpha, n]/normg[alpha]
qppq[kappa_]:=N[Sqrt[Sum[Abs[Sum[Sum[Exp[2 Pi I m (n-l)/d] m (n-l)
   psi[kappa,l]/d,{l,-s,s}],{m,-s,s}]-I d psi[kappa,n]/(2 Pi)]^2,{n,-s,s}]]]
Plot[qppq[x], {x, 1/2, 2}, AxesLabel -> {k,}]
\end{verbatim}
\rule{2mm}{2mm}\ \ {\bf {\small FIGURE} 3}\\[2mm]
Fig.3 has been obtained by using for $d\!=\!11$ and $d\!=\!31$ the program:\\[-5mm]
\begin{verbatim}
d = 11; s := (d - 1)/2 
g[alpha_, n_] := N[Sum[Exp[-Pi  alpha (m d + n)^2/d], {m, -Infinity, Infinity}]]
normg[alpha_] := N[Sqrt[Sum[g[alpha, m]^2, {m, -s, s}]]]
mod11[kappa_]:= N[Abs[Abs[(1/(normg[kappa]^2)) Sum[g[kappa,n] Sum[Sum[Exp[2 Pi I m (n-l)/d]
   m(n-l)g[kappa,l]/d,{l,-s,s}],{m,-s,s}],{n,-s,s}]]-d/(2 Pi)]]
Plot[mod11[x], {x, 1/4, 4}, AxesLabel -> {k,}]
\end{verbatim}
\rule{2mm}{2mm}\ \ {\bf {\small FIGURE} 4}\\[2mm]
Fig.4 has been obtained by using   the program:\\[-5mm]
\begin{verbatim}
d = 31; s := (d - 1)/2
HOscillator[n_, m_] :=  N[(1/2) n^2 DiscreteDelta[n - m] 
   + (1/(2 d)) Sum[k^2 Exp[2 Pi I k (n - m)/d], {k, -s, s}]]
Osc := Table[HOscillator[n, m], {n, -s, s}, {m, -s, s}]
hh[n_] := Re[Eigenvalues[Osc][[n]]]
ListPlot[Table[{n - 1, hh[d - n + 1]}, {n, 1, d}], Filling -> Axis, AxesLabel -> {n,}, 
   AspectRatio -> 2/3, PlotRange -> Full]
\end{verbatim}
\rule{2mm}{2mm}\ \ {\bf {\small FIGURE} 5}\\[2mm]
Fig.5 has been obtained by using  the program:\\[-5mm]
\begin{verbatim}
d = 31; s := (d - 1)/2
v[k_] := N[Sum[Exp[-Pi (a d + k)^2/d], {a, -10, 10}]]
w[k_] := N[v[k]/Sqrt[Sum[v[n]^2, {n, 0, d - 1}]]]
QOscillator[n_, k_] := (1/(2 d)) Sum[(a^2+b^2) Exp[2 Pi I b (n-k)/d] w[a+n] w[a+k],{a,-s,s},{b,-s,s}]
QOsc := Table[QOscillator[n, k], {n, -s, s}, {k, -s, s}]
ReEigVal = Re[Eigenvalues[QOsc]]
hh[n_] := ReEigVal[[n]]
ListPlot[Table[{n - 1, ReEigVal[[d - n + 1]]}, {n, 1, d}],  Filling -> Axis, AxesLabel -> {n,}, 
    AspectRatio -> 2/3,  PlotRange -> Full]
\end{verbatim}
\rule{2mm}{2mm}\ \ {\bf {\small FIGURE} 6}\\[2mm]
Fig.6 has been obtained by using  the program:\\[-5mm]
\begin{verbatim}
d = 31; s := (d - 1)/2
HOscillator[n_, m_]:=-N[2 (N[Cos[2 Pi n/d]]-2) DiscreteDelta[n-m]+ DiscreteDelta[n-m-1]
    +DiscreteDelta[n-m+1]+DiscreteDelta[n-m-2 s]+DiscreteDelta[n-m+2 s]]
HOsc := Table[HOscillator[n, m], {n, -s, s}, {m, -s, s}]
hh[n_] := Re[Eigenvalues[HOsc][[n]]]
ListPlot[Table[{n - 1, hh[d - n + 1]}, {n, 1, d}], Filling -> Axis, AxesLabel -> {n,}, 
   AspectRatio -> 2/3, PlotRange -> Full]
\end{verbatim}
\rule{2mm}{2mm}\ \ {\bf {\small TABLE} I}\\[2mm]
Table I has been obtained by using for $d\!=\!11$ and $d\!=\!31$ the program:\\[-5mm]
\begin{verbatim}
d = 11; s = (d - 1)/2
QQ := Table[n DiscreteDelta[Mod[n - k, d]], {n, -s, s}, {k, -s, s}]
FF := Table[N[(1/Sqrt[d]) Exp[-2 Pi I n k/d]], {n, -s, s}, {k, -s, s}]
PP := ConjugateTranspose[FF] . QQ. FF
DD := Table[DiscreteDelta[n - k], {n, -s, s}, {k, -s, s}]
EE := QQ.PP - PP.QQ - I (d/(2 Pi)) DD
Reverse[Im[Eigenvalues[EE]]]
\end{verbatim}
\rule{2mm}{2mm}\ \ {\bf {\small TABLE} II}\\[2mm]
Table II has been obtained by using the program:\\[-5mm]
\begin{verbatim}
d = 11; s := (d - 1)/2 
QQ := Table[j DiscreteDelta[Mod[j - k, d]], {j, -s, s}, {k, -s, s}]
FF := Table[N[(1/Sqrt[d]) Exp[-2 Pi I j k/d]], {j, -s, s}, {k, -s, s}]
PP := ConjugateTranspose[FF] . QQ. FF
DD := Table[DiscreteDelta[j - k], {j, -s, s}, {k, -s, s}]
EE := QQ.PP - PP.QQ - I (d/(2 Pi)) DD
phi[k_] := Eigenvectors[EE][[d - k + 1]]
Table[Norm[FF.phi[kk] - phi[kk]], {kk, 1, d}]
Table[Norm[FF.phi[kk] + phi[kk]], {kk, 1, d}]
Table[Norm[FF.phi[kk] - I phi[kk]], {kk, 1, d}]
Table[Norm[FF.phi[kk] + I phi[kk]], {kk, 1, d}]
\end{verbatim}
\rule{2mm}{2mm}\ \ {\bf {\small TABLE} III}\\[2mm]
Table III has been obtained by using for $\kappa\!=\!1,\, \kappa\!=\!2,\, \kappa\!=\!1/2,\, \kappa\!=\!3\ \mbox{and}\ \kappa\!=\!1/3$ the program:\\[-5mm]
\begin{verbatim}
kappa = 1; d = 11; s := (d - 1)/2 
g[alpha_, n_] := N[Sum[Exp[-Pi  alpha (m d + n)^2/d], {m, -Infinity, Infinity}]]
normg[alpha_] := N[Sqrt[Sum[g[alpha, m]^2, {m, -s, s}]]]
gamma := Table[ g[kappa, n]/normg[kappa], {n, -s, s}]
QQ := Table[j DiscreteDelta[Mod[j - k, d]], {j, -s, s}, {k, -s, s}]
FF := Table[N[(1/Sqrt[d]) Exp[-2 Pi I j k/d]], {j, -s, s}, {k, -s, s}]
PP := ConjugateTranspose[FF] . QQ. FF
DD := Table[DiscreteDelta[j - k], {j, -s, s}, {k, -s, s}]
EE := QQ.PP - PP.QQ - I (d/(2 Pi)) DD
phi[k_] := Eigenvectors[EE][[d - k + 1]]
N[Table[Abs[gamma .Transpose[{phi[k]}]], {k, 1, d}]]
\end{verbatim}
\rule{2mm}{2mm}\ \ {\bf {\small TABLE} IV }\\[2mm]
Table IV has been obtained by using for $d\!=\!11,\, d\!=\!31$ and $\kappa\!=\!1,\, \kappa\!=\!2,\, ...\, \kappa\!=\!1/3$ the program:\\[-5mm] 
\begin{verbatim}
kappa=1; d = 11; s:=(d-1)/2 
g[alpha_, n_] :=  N[Sum[Exp[-Pi  alpha (m d + n)^2/d], {m, -Infinity, Infinity}]]
normg[alpha_] := N[Sqrt[Sum[g[alpha, m]^2, {m, -s, s}]]]
N[(1/normg[kappa]) Sqrt[Sum[Abs[Sum[Sum[Exp[2 Pi I m (n-l)/d] m (n-l) g[kappa,l]/d, 
   {l,-s,s}], {m,-s,s}]-I d g[kappa,n]/(2 Pi)]^2,{n,-s,s}]]]
\end{verbatim}
\rule{2mm}{2mm}\ \ {\bf {\small TABLE} V }\\[2mm]
Table V has been obtained by using for $d\!=\!11,\, d\!=\!31$ and $\kappa\!=\!1,\, \kappa\!=\!2,\, ...\, \kappa\!=\!1/8$ the program:\\[-5mm] 
\begin{verbatim}
kappa=1; d=11; s:=(d-1)/2 
g[alpha_, n_] :=  N[Sum[Exp[-Pi  alpha (m d + n)^2/d], {m, -Infinity, Infinity}]]
normg[alpha_] := N[Sqrt[Sum[g[alpha, m]^2, {m, -s, s}]]]
N[Abs[(1/ (normg[kappa]^2)) Sum[g[kappa, n] Sum[Sum[Exp[2 Pi I m (n-l) /d] m (n-l) 
   g[kappa, l]/d,{l,-s,s}],{m,-s,s}],{n,-s,s}]]-d/(2 Pi)]
\end{verbatim}
\rule{2mm}{2mm}\ \ {\bf {\small TABLE} VI }\\[2mm]
Table VI has been obtained by using for $k\!=\!0$, $k\!=\!1$, ... , $k\!=\!10$ the program:\\[-5mm] 
\begin{verbatim}
k = 2; d = 31; s := (d - 1)/2
g[n_] := N[Sum[Exp[-Pi (m d + n)^2/d], {m, -Infinity, Infinity}]]
normg := N[Sqrt[Sum[g[m]^2, {m, -s, s}]]]
gg := Table[g[n]/normg, {n, -s, s}]
QQ := Table[j DiscreteDelta[Mod[j - m, d]], {j, -s, s}, {m, -s, s}]
FF := Table[N[(1/Sqrt[d]) Exp[-2 Pi I j m/d]], {j, -s, s}, {m, -s, s}]
PP := ConjugateTranspose[FF].QQ.FF
CC := Sqrt[Pi/d] (QQ - I PP)
phi := MatrixPower[CC, k].gg
psi = phi/Norm[phi]
N[Sqrt[Sum[Abs[Sum[Sum[Exp[2 Pi I m (j-l)/d] m (j-l) psi[[l+s+1]]/d,{l,-s,s}],{m,-s,s}]- 
      I d psi[[j + s + 1]]/(2 Pi)]^2, {j, -s, s}]]]
N[Abs[Abs[Sum[Conjugate[psi[[j+s+1]]]Sum[Sum[Exp[2 Pi I m (j-l)/d] m (j-l)psi[[l+s+1]]/d,
    {l, -s, s}], {m, -s, s}], {j, -s, s}]] - d /(2 Pi)]]
\end{verbatim}
\rule{2mm}{2mm}\ \ {\bf {\small TABLE} VII }\\[2mm]
We firstly obtain the matrix EigVect containing as rows the eigenvectors of $\mathfrak{H}$ by using the program:\\[-5mm] 
\begin{verbatim}
d = 31; s := (d - 1)/2
HOscillator[n_, m_] :=  N[(1/2) n^2 DiscreteDelta[n - m] 
   + (1/(2 d)) Sum[k^2 Exp[2 Pi I k (n - m)/d], {k, -s, s}]]
Osc := Table[HOscillator[n, m], {n, -s, s}, {m, -s, s}]
EigVect = Reverse[Re[Eigenvectors[Osc]]]
\end{verbatim}
Next, we Copy/Paste the matrix EigVect in the following program and consider $k\!=\!0$, $k\!=\!1$, ..., $k\!=\!10$, 
\begin{verbatim}
k=0;  d := 31; s := (d - 1)/2
EigVect := ..............
EigVk:=EigVect[[k]]
N[Sqrt[Sum[Abs[Sum[Sum[Exp[2 Pi I m (j - l)/d] m (j - l) EigVk[[l + s + 1]]/
          d, {l, -s, s}], {m, -s, s}] - I d EigVk[[j + s + 1]]/(2 Pi)]^2, {j, -s, s}]]]
N[Abs[Abs[Sum[Conjugate[EigVk[[j + s + 1]]] Sum[ Sum[Exp[2 Pi I m (j - l)/d] 
   m (j - l) EigVk[[l + s + 1]]/ d, {l, -s, s}], {m, -s, s}], {j, -s, s}]] - d/(2 Pi)]]
\end{verbatim}
\rule{2mm}{2mm}\ \ {\bf {\small TABLE} VIII }\\[2mm]
Table VIII has been obtained by using for $n\!=\!0$, $n\!=\!1$, ... , $n\!=\!5$ and $k\!=\!0$, $k\!=\!1$, ... , $k\!=\!5$ the program:\\[-5mm] 
\begin{verbatim}
n = 0; k = 1; d = 31; s := (d - 1)/2 
g[m_] := N[Sum[Exp[-Pi (a d + m)^2/d], {a, -10, 10}]]
phi[m_] := N[Exp[-Pi I n k/d] Exp[2 Pi I k m/d] g[m - n]]
nphi := N[Sqrt[Sum[Abs[phi[m]]^2, {m, -s, s}]]]
psi[m_] := N[(1/nphi) phi[m]]
N[Sqrt[Sum[Abs[Sum[Sum[Exp[2 Pi I m (j-l) /d] m (j-l) psi[l]/d,{l,-s, s}], 
   {m,-s,s}]-I d psi[j]/(2 Pi)]^2,{j,-s,s}]]]
N[Abs[Abs[Sum[Conjugate[psi[j]] Sum[Sum[Exp[2 Pi I m (j-l)/d] m (j-l) psi[l]/d,
   {l,-s, s}],{m,-s,s}],{j,-s,s}]]-d /(2 Pi)]]
\end{verbatim}
\rule{2mm}{2mm}\ \ {\bf {\small TABLE} IX}\\[2mm]
We firstly obtain the matrix EigVect containing as rows the eigenvectors of $\mathfrak{a}$ by using the program:\\[-5mm] 
\begin{verbatim}
d = 31; s := (d - 1)/2
Annihilation[n_, m_]:= N[n DiscreteDelta[n-m]+I/d Sum[k Exp[2 Pi I k (n-m)/d],{k,-s,s}]]
Annih := Table[Annihilation[n, m], {n, -s, s}, {m, -s, s}]
EigVec = Reverse[Eigenvectors[Annih]]
\end{verbatim}
Next, we Copy/Paste the matrix EigVect in the following program and consider $k\!=\!0$, $k\!=\!1$, ..., $k\!=\!19$, 
\begin{verbatim}
k=0;  d := 31; s := (d - 1)/2
EigVect := ..............
EigVk:=EigVect[[k+1]]
N[Sqrt[Sum[Abs[Sum[Sum[Exp[2 Pi I m (j - l)/d] m (j - l) EigVk[[l + s + 1]]/
          d, {l, -s, s}], {m, -s, s}] - I d EigVk[[j + s + 1]]/(2 Pi)]^2, {j, -s, s}]]]
N[Abs[Abs[Sum[Conjugate[EigVk[[j + s + 1]]] Sum[ Sum[Exp[2 Pi I m (j - l)/d] 
   m (j - l) EigVk[[l + s + 1]]/ d, {l, -s, s}], {m, -s, s}], {j, -s, s}]] - d/(2 Pi)]]
\end{verbatim}
\rule{2mm}{2mm}\ \ {\bf {\small TABLE} X }\\[2mm]
We firstly obtain the matrix EigVect containing as rows the eigenvectors of $\mathfrak{H}_q$ by using the program:\\[-5mm] 
\begin{verbatim}
d = 31; s := (d - 1)/2
v[k_] := N[Sum[Exp[-Pi (a d + k)^2/d], {a, -10, 10}]]
w[k_] := N[v[k]/Sqrt[Sum[v[n]^2, {n, 0, d - 1}]]]
QOscillator[n_, k_] := (1/(2 d)) Sum[(a^2+b^2) Exp[2 Pi I b (n-k)/d] w[a+n] w[a+k],{a,-s,s},{b,-s,s}]
QOsc := Table[QOscillator[n, k], {n, -s, s}, {k, -s, s}]
EigVec = Reverse[Re[Eigenvectors[QOsc]]]
\end{verbatim}
Next, we Copy/Paste the matrix EigVect in the following program and consider $k\!=\!0$, $k\!=\!1$, ..., $k\!=\!10$, 
\begin{verbatim}
k=0;  d := 31; s := (d - 1)/2
EigVect := ..............
EigVk:=EigVect[[k+1]]
N[Sqrt[Sum[Abs[Sum[Sum[Exp[2 Pi I m (j - l)/d] m (j - l) EigVk[[l + s + 1]]/
          d, {l, -s, s}], {m, -s, s}] - I d EigVk[[j + s + 1]]/(2 Pi)]^2, {j, -s, s}]]]
N[Abs[Abs[Sum[Conjugate[EigVk[[j + s + 1]]] Sum[ Sum[Exp[2 Pi I m (j - l)/d] 
   m (j - l) EigVk[[l + s + 1]]/ d, {l, -s, s}], {m, -s, s}], {j, -s, s}]] - d/(2 Pi)]]
\end{verbatim}
\rule{2mm}{2mm}\ \ {\bf {\small TABLE} XI }\\[2mm]
Table XI has been obtained by using for $k\!=\!0$, $k\!=\!1$, ... , $k\!=\!10$ the program:\\[-5mm] 
\begin{verbatim}
k = 0; d = 31; s := (d - 1)/2 
phi[n_] :=N[Sum[HermiteH[k, a d + n] Exp[-Pi (a d + n)^2/d], {a, -10, 10}]]
nphi := N[Sqrt[Sum[phi[n]^2, {n, -s, s}]]]
psik = N[Table[ phi[n]/nphi, {n, -s, s}]]
N[Sqrt[Sum[Abs[Sum[Sum[Exp[2 Pi I m (n-l)/d] m (n-l) psik[[l+s+1]]/d,
   {l,-s,s}], {m,-s,s}]- I d psik[[n+s+1]]/(2 Pi)]^2,{n,-s,s}]]]
N[Abs[Abs[Sum[Conjugate[psik[[j+s+1]]] Sum[Sum[Exp[2 Pi I m(j-l)/d]m(j-l)
    psik[[l+s+1]]/d,{l,-s,s}],{m,-s,s}],{j,-s,s}]]- d/(2 Pi)]]
\end{verbatim}
\rule{2mm}{2mm}\ \ {\bf {\small TABLE} XII}\\[2mm]
We firstly obtain the matrix EigVect containing as rows the eigenvectors of $\mathfrak{a}$ by using the program:\\[-5mm] 
\begin{verbatim}
d = 31; s := (d - 1)/2
HOscillator[n_, m_]:=-N[2 (N[Cos[2 Pi n/d]]-2) DiscreteDelta[n-m]+ DiscreteDelta[n-m-1]
    +DiscreteDelta[n-m+1]+DiscreteDelta[n-m-2 s]+DiscreteDelta[n-m+2 s]]
HOsc := Table[HOscillator[n, m], {n, -s, s}, {m, -s, s}]
EigVect = Reverse[Eigenvectors[HOsc]]
\end{verbatim}
Next, we Copy/Paste the matrix EigVect in the following program and consider $k\!=\!0$, $k\!=\!1$, ..., $k\!=\!19$, 
\begin{verbatim}
k=0;  d := 31; s := (d - 1)/2
EigVect := ..............
EigVk:=EigVect[[k+1]]
N[Sqrt[Sum[Abs[Sum[Sum[Exp[2 Pi I m (j - l)/d] m (j - l) EigVk[[l + s + 1]]/
          d, {l, -s, s}], {m, -s, s}] - I d EigVk[[j + s + 1]]/(2 Pi)]^2, {j, -s, s}]]]
N[Abs[Abs[Sum[Conjugate[EigVk[[j + s + 1]]] Sum[ Sum[Exp[2 Pi I m (j - l)/d] 
   m (j - l) EigVk[[l + s + 1]]/ d, {l, -s, s}], {m, -s, s}], {j, -s, s}]] - d/(2 Pi)]]
\end{verbatim}
\rule{2mm}{2mm}\ \ {\bf {\small TABLE} XIII }\\[2mm]
Table XIII has been obtained by using for $k\!=\!0$, $k\!=\!1$, $k\!=\!3$ the program:\\[-5mm] 
\begin{verbatim}
k = 0; d = 31; s := (d - 1)/2 
K[m_, l_] := (1/2^s) Sqrt[Binomial[2 s,s+l]/ Binomial[2 s, s+m]] Sum[(-1)^n Binomial[s+l,n]
    Binomial[s-1l,s+m-n],{n,0,s+m}]
KK := N[Table[ K[k - s, n], {n, -s, s}]]
N[Sqrt[Sum[Abs[Sum[Sum[Exp[2 Pi I m (n-l)/d] m(n-l) KK[[l+j+1]]/d,{l,-s,s}],{m,-s,s}]
   -I d KK[[n+j+1]]/(2 Pi)]^2,{n,-s,s}]]]
N[Abs[Abs[Sum[Conjugate[KK[[j+s+1]]] Sum[Sum[Exp[2 Pi I m(j-l)/d] m(j-l) KK[[l+s+1]]/d, 
   {l,-s,s}],{m,-s,s}],{j,-s,s}]]-d/(2 Pi)]]
\end{verbatim}

\end{document}